\documentclass[a4paper,11pt]{article}
\usepackage{jheppub} 

\usepackage[normalem]{ulem}

\usepackage{float}
\usepackage{mmacells}
\usepackage{caption}
\usepackage{subcaption}
 \allowdisplaybreaks
\newcommand{\mt}{\textsc{Mathematica}\,}
\newcommand{\ar}{\texttt{AlgRel.wl}\,}
\newcommand{\eg}{\texttt{Examples.nb}\,}

\title{\texttt{AlgRel.wl}: Algebraic relations for the product of propagators in Feynman integrals}
\author[]{B. Ananthanarayan,}
\author[]{Souvik Bera,}
\author[]{Tanay Pathak}

\affiliation[a]{Centre for High Energy Physics, Indian Institute of Science,\\ Bangalore-560012, Karnataka, India}

\emailAdd{anant@iisc.ac.in, souvikbera@iisc.ac.in, tanaypathak@iisc.ac.in}
\abstract{
 Motivated by the foundational work of Tarasov, who pointed out that the algebraic relations of the type considered here can lead to functional reduction of Feynman integrals, we suitably modify the original method to be able to implement and automatize it and present a \textsc{Mathematica} package \texttt{AlgRel.wl}. The purpose of this package is to help derive the algebraic relations with arbitrary kinematic quantities, for the product of propagators. Under specific choices of the arbitrary parameters that appear in these relations, we can write the original integral with all massive propagators in general, as a sum of integrals which have fewer massive propagators. The resulting integrals are of reduced complexity for computational purposes. For the one-loop cases, with all different and non-zero masses, this would result in integrals with one massive propagator. We also devise a strategy so that the method can also be applied to higher-loop integrals. We demonstrate the procedure and the results obtained using the package for various one-loop and higher-loop examples.  Due to the fact that the Feynman integrals are intimately related to the hypergeometric functions, a useful consequence of these algebraic relations is in deriving the sets of non-trivial reduction formulae. We present various such reduction formulae and further discuss how, more such formulae can be obtained than described here. The \ar package and an example notebook \eg can be found at \href{https://github.com/TanayPathak-17/Algebraic-relation-for-the-product-of-propagators}{GitHub}.}

\date{Centre for High Energy Physics, Indian Institute of Science,\\ Bangalore-560012, Karnataka, India
}
\begin{document}
\maketitle

\section{Introduction}
In this work, we consider the formalism first proposed by Tarasov to derive algebraic relations for the product of propagators for functional reduction \cite{Tarasov:2015wcd}.  We systematically develop an algorithm inspired by the original work and
present a realization in \mt for the same, which is provided for the user as a package called \ar.
We have used the package to simplify and analyze many important and interesting Feynman Integrals that are
amenable to treatment using this formalism. 

Feynman integrals play an important role in precision calculations in quantum field theory. There are various methods to evaluate them \cite{smirnov2006feynman,Weinzierl:2022eaz}. Even with all these methods, it is at times still challenging to compute Feynman integrals. More often, other techniques are used to facilitate this computation.  In \cite{Tarasov:2008hw} the method of functional reduction is introduced to derive functional relations between Feynman integrals. These relations reduce the original integral into a sum of integrals which are easier to evaluate. The focus of the present work is this new way to obtain functional relations by deriving the algebraic relations for the product of propagators. This method in turn then leaves some undetermined free parameters which can be chosen at will. Appropriate choices of these parameters result in various functional equations for Feynman integrals\cite{Tarasov:2022pwt,Tarasov:2019tgs,Tarasov:2019mqy,Tarasov:2017yyd,Kniehl:2016yrh}.

The method can be applied to any one-loop diagram, indeed as already pointed out in
detail by Tarasov.  Despite this, no working code has been provided in the past. In the present work, we provide an automated \mt package \ar to derive the algebraic relation for the product of propagators. Our code here fills this gap in the possibility of finding widespread use of formalism. Since our goal is an efficient algorithmic implementation to find the algebraic relation, we introduce a recursive way of method. The free parameters in the resulting relation can then be chosen in an appropriate manner to derive the functional equations for the Feynman integrals. More specifically, for presentation purposes, we focus on the cases when all these free parameters are zero and the original Feynman integral with many massive propagators can be written as a sum of integrals with fewer massive propagators, which was also pointed out in \cite{Tarasov:2011zz}\footnote{We also briefly discuss a case when we choose a non-zero parameter in Appendix \ref{append:funcred}}.  For the one-loop integrals, with all different and non-zero masses, this procedure can be used to reduce the original integral to a sum of integrals with one massive propagator. We apply the method for up to 6-point, one-loop integrals and show that the $N-$point one loop integral with all massive propagators and general external momenta can be written as a sum of $2^{N-1}$ integrals with just one massive propagator. Though the method is not readily generalizable to the higher loop we yet extend the uses to cover
certain cases of 2- and even 3-loop Feynman integrals. In a similar manner, this approach is also applicable to higher loops. Our findings show that we require at least
4 propagators in order for the formalism to be viable and to be of utility as far as the simplification is
concerned. We explain this feature in some detail. 

We, however, notice that such functional reduction is one of the many possibilities obtained after choosing the free parameters obtained from the algebraic relation \cite{Tarasov:2022clb}. 
In view of the proposed method of functional reduction of Feynman integrals, the package has been built in such a way that the final result still has arbitrary parameters which can be chosen suitably for the functional reduction procedure. Using a few of the analytical results available for the one-loop integrals, we explicitly show how the complexity in the evaluation of these integrals can be reduced. Whenever the Feynman integrals can be expressed in terms of hypergeometric functions \cite{delaCruz:2019skx,Klausen:2019hrg,Ananthanarayan:2022ntm,Blumlein:2021hbq} this reduction in complexity gives rise to reduction formulae for the hypergeometric functions. Thus it can be used to establish new relations between multi-variable hypergeometric functions. We discover many new reduction formulae for such hypergeometric functions, which, to the
best of our knowledge,  have not appeared anywhere in the literature. We also discuss in detail how further reduction formulae can be obtained from already available results for the one-loop cases. Such relations between hypergeometric functions are also obtained in \cite{Kniehl:2011ym}, where explicit relations between hypergeometric functions are derived via the evaluation of Feynman integrals. In order to make the results accessible, we provide several examples in a single \mt notebook that allows
the reader to appreciate the power of the formalism, based on the code that is provided along with it.

The article is organized as follows. In section \ref{secMethod} we discuss the method in detail with one loop bubble integral as an example and explicitly present how the reduction in complexity has been achieved for the integral. In section \ref{algorithm}, we  present the algorithm of the \ar package and discuss its usage in detail. In section \ref{results}, various results obtained for one, two, and three-loop integrals are presented. In section \ref{hypergeometric}, we  discuss the various analytic results in terms of multi-variable hypergeometric functions already derived for the one-loop $N-$point integrals \cite{davydychev1991some,davydychev1992general} and show how the present work helps in deriving the reduction formulae for the multi-variable hypergeometric functions using them. Finally, we conclude the paper with summary and discussions in section \ref{conclusion}. In Appendix \ref{appendix:redfor}, we provide a list of various reduction formulae that we derive, along with some details on how to further extend the list given there.

The package \ar along with a \mt notebook \eg, that contains all the examples discussed in the paper can be found in the \href{https://github.com/TanayPathak-17/Algebraic-relation-for-the-product-of-propagators}{GitHub} repository.

\section{The Method}\label{secMethod}
We now explain the method to find the algebraic relation of the product of propagators with the help of the one-loop bubble integral. For this example and all the following examples we consider the integrals in $d$ dimensions.
\begin{figure}[htbp]
\centering\includegraphics[width=.3\textwidth]{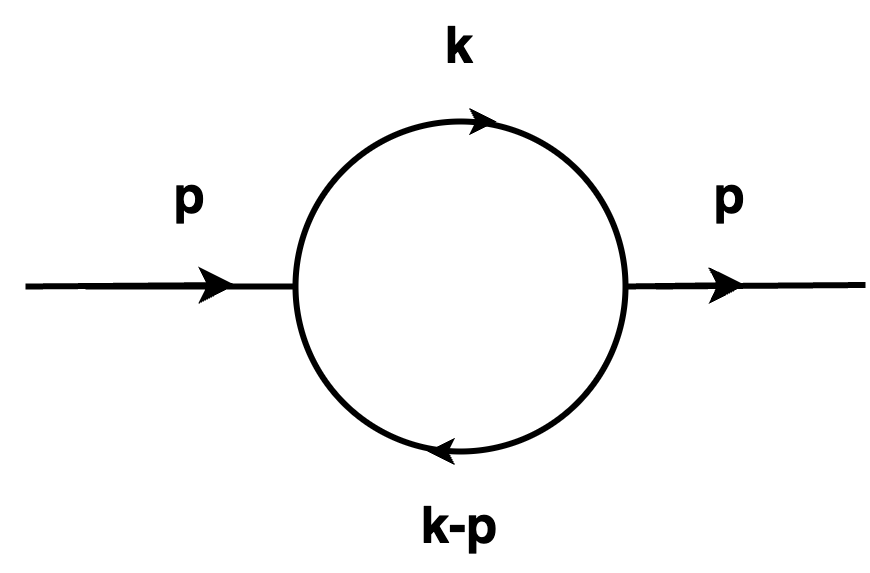}
\caption{Bubble diagram}\label{figbubble}
\end{figure}
Consider the one-loop bubble integral corresponding to bubble diagram Fig.\ref{figbubble},
\begin{equation}\label{int2point}
    I_{2}(p^{2},m_1,m_2)= \int \frac{d^{d}k}{(k^{2}-m_{1}^{2})((k-p)^{2}-m_{2}^{2})}
\end{equation}
To find the algebraic relation for the product of propagators, we instead consider a more general propagator, depending on only one loop-momenta, of the following form
\begin{align}\label{eqpropgen}
    d_{i}= (k+q_{i})^{2}-m_{i}^{2}
\end{align}
where $k$ is the loop-momentum, $q_{i}$'s are dependent on external momenta and can be zero as well and $m_{i}$ is the mass of the propagator.

\begin{figure}[h]
\centering\includegraphics[width=.85\textwidth]{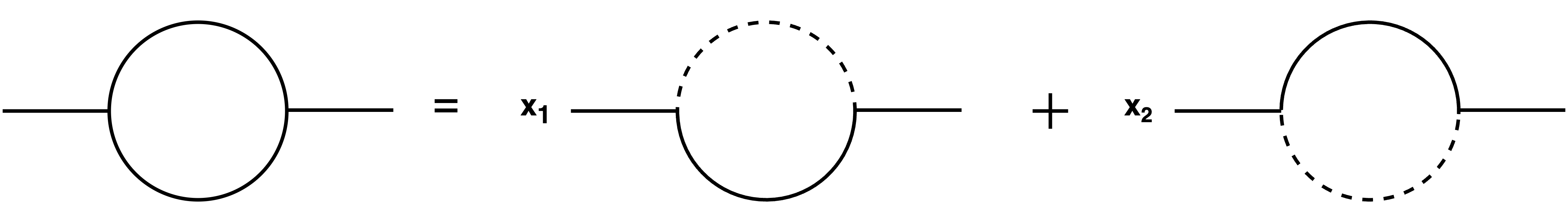}
\caption{Diagrammatic representation of Eq.\eqref{redeqbubble}}\label{figbubblediag}
\end{figure}

With the general propagators, we now have
\begin{equation}
     I_{2}((q_1-q_2)^{2},m_1,m_2) = \int \frac{d^{d}k}{d_{1}d_{2}}
\end{equation}
where substituting $q_{1}=0$ and $q_{2}=-p$ we recover Eq.\eqref{int2point}.\\
We seek the algebraic relation for the integrand, by introducing a new denominator $D_{1}$ along with coefficients $x_1$ and $x_2$, of the following form
\begin{equation}\label{integranbubble}
    \frac{1}{d_{1}d_{2}} = \frac{x_{1}}{D_{1} d_{1}} + \frac{x_{2}}{D_{1} d_{2}}
\end{equation}
where $D_{i}= (k+P_{i})^{2}-M_{i}^{2}$ is defined similar to Eq.\eqref{eqpropgen}.\\
The unknowns that are introduced can be fixed using the above equation, while the remaining parameters are arbitrary and can be fixed at will in such a way that the resulting relationship gives rise to integrals which are easier to compute.\\ 
Using Eq.\eqref{integranbubble} we get
\begin{equation}
    D_{1}= x_{1} d_{2} +x_{2} d_{1}
\end{equation}
Comparing the coefficients of $k^{2}$, $k$ and the remaining $k$ independent term we get 
\begin{align}
    &x_1+x_2 =1 \nonumber \\
    & x_1 \mathbf{q_2} + x_2 \mathbf{q_1} = \mathbf{P_1} \nonumber \\
   &-M_{1}^2 + P_{1}^2 - (-m_{2}^2 + q_{2}^2) x_{1} - (-m_{1}^2 + q_{1}^2) x_2 =0 
\end{align}
Solving for $x_1$, $x_2$ and $\mathbf{P_{1}}$ we get following two sets of solutions
\begin{align}\label{bubxxp}
   x_{1}&= \frac{\sqrt{\left(m_1^2-m_2^2+\left(q_1-q_2\right){}^2\right){}^2-4 \left(q_1-q_2\right){}^2 \left(m_1^2-M_1^2\right)}+m_1^2-m_2^2+q_1^2+q_2^2-2 q_1 q_2}{2 \left(q_1-q_2\right){}^2} \nonumber \\
   x_{2} &= \frac{-\sqrt{\left(m_1^2-m_2^2+\left(q_1-q_2\right){}^2\right){}^2-4 \left(q_1-q_2\right){}^2 \left(m_1^2-M_1^2\right)}-m_1^2+m_2^2+q_1^2+q_2^2-2 q_1 q_2}{2 \left(q_1-q_2\right){}^2} \nonumber \\
   \mathbf{P_1}&= \frac{(\mathbf{q_1-q_2})(-\sqrt{(m_1^2-m_2^2+(q_1-q_2){}^2){}^2-4 (q_1-q_2){}^2 (m_1^2-M_1^2)}-m_1^2+m_2^2+q_1^2+q_2^2-2 q_1 q_2)}{2 (q_1-q_2){}^2}+\mathbf{q_2}
\end{align}
and 
\begin{align}\label{bubxxp2}
   x_{1}&= \frac{-\sqrt{\left(m_1^2-m_2^2+\left(q_1-q_2\right){}^2\right){}^2-4 \left(q_1-q_2\right){}^2 \left(m_1^2-M_1^2\right)}+m_1^2-m_2^2+q_1^2+q_2^2-2 q_1 q_2}{2 \left(q_1-q_2\right){}^2} \nonumber \\
   x_{2} &= \frac{\sqrt{\left(m_1^2-m_2^2+\left(q_1-q_2\right){}^2\right){}^2-4 \left(q_1-q_2\right){}^2 \left(m_1^2-M_1^2\right)}-m_1^2+m_2^2+q_1^2+q_2^2-2 q_1 q_2}{2 \left(q_1-q_2\right){}^2} \nonumber \\
   \mathbf{P_1}&=\frac{\mathbf{(q_1-q_2)} (\sqrt{(m_1^2-m_2^2+(q_1-q_2){}^2){}^2-4 (q_1-q_2){}^2 (m_1^2-M_1^2)}-m_1^2+m_2^2+q_1^2+q_2^2-2 q_1 q_2)}{2 (q_1-q_2){}^2}+\mathbf{q_2}
\end{align}
We remark that both of the above sets can be used for the purpose of finding the algebraic relation. However, for convenience, we would use the set given by Eq.\eqref{bubxxp}. For the purpose of the algebraic relations, both of the choices are equivalent. One can use either of the solutions and can check that it satisfies the algebraic relation that we seek to find.
In the above equation, $M_1$ is still an arbitrary variable that can be chosen at will. Choosing various values of $M_{1}$ results in different functional equations \cite{Tarasov:2022clb} for the bubble integral. For the present work, we focus on one of the simple choices i.e. $M_{1}=0$. Integrating Eq.\eqref{integranbubble} and substituting $q_{1}=0$ and $q_{2}= -p$ we have
\begin{equation}\label{redeqbubble}
    I_{2}(p^{2},m_{1},m_{2})= x_{1} I_{2}((P_1+p)^{2},0,m_{2})+ x_{2} I_{2}(P_{1}^{2},m_{1},0)
\end{equation}
Hence we see that the general two-point bubble integral with non-zero masses can be written in terms of two integrals with just one mass. Diagrammatically Eq.\eqref{redeqbubble} can be represented as in Fig.\ref{figbubblediag}.

\noindent To see how the complexity in the computation has been reduced in the Eq.\eqref{redeqbubble}, we refer to a few analytic results. The general result for the massive bubble diagram can be written in terms of the Appell $F_4$ function \cite{GONZALEZ201050}. 
\begin{align}\label{bubf4}
  I_{2}(p,m_{1},m_{2})&=  \frac{(m_2^2){}^{\frac{d}{2}-2} \Gamma (\frac{d}{2}-1) \Gamma (2-\frac{d}{2})}{\Gamma (\frac{d}{2})} F_{4}\Big(2-\frac{d}{2},1;\frac{d}{2},2-\frac{d}{2};\frac{p^2}{m_2^2},\frac{m_1^2}{m_2^2}\Big)\nonumber \\
  &+\frac{(m_1^2){}^{\frac{d}{2}-1} \Gamma (1-\frac{d}{2}) }{m_2^2} F_{4}\Big(\frac{d}{2},1;\frac{d}{2},\frac{d}{2};\frac{p^2}{m_2^2},\frac{m_1^2}{m_2^2}\Big)
\end{align}
where, 
\begin{equation}
    F_4(a,b,c,d,x,y)= \sum_{m,n=0}^{\infty}\dfrac{(a)_{m+n}(b)_{m+n}}{(c)_{m}(d)_{n}m!n!}x^{m}y^{n} 
\end{equation} is the Appell $F_{4}$ hypergeometric series with region of convergence (ROC) given by $\sqrt{|x|}+ \sqrt{|y|} < 1$.\\
The analytic expression result for $I_{2}(p,m,0)$ is readily available in  \cite{bollini1972lowest, Boos:1990rg}.
\begin{equation}
    I_2^{(d)}( p^2; m^2, 0 )=-\Gamma(1-\frac{d}{2}) m_2^{d-4} \,_{2}F_{1}\left[\begin{array}{c}1,2-\frac{d}{2} ; \\ \frac{d}{2} ;\end{array} \frac{p^2}{m^2}\right]
\end{equation}
Using the above relation in Eq.\eqref{redeqbubble}, we get the following for the right-hand side
\begin{align}\label{bub2f1}
    &-\frac{m_1^{d-4} \Gamma \left(1-\frac{d}{2}\right) \left(\sqrt{\left(-m_1^2+m_2^2+p^2\right){}^2-4 m_1^2 p^2}+m_1^2-m_2^2+p^2\right)}{2 p^2}\nonumber \\
   &\times \,_{2}F_{1}\left[\begin{array}{c}1,2-\frac{d}{2} ; \\ \frac{d}{2} ;\end{array} \frac{\left(p^2+m_1^2-m_2^2+\sqrt{\left(p^2+m_1^2-m_2^2\right){}^2-4 p^2 m_1^2}\right){}^2}{4 p^2 m_1^2}\right]\nonumber\\
   &-\frac{m_2^{d-4} \Gamma \left(1-\frac{d}{2}\right)}{2 p^2} (m_1^2-m_2^2+p^2+\sqrt{(-m_1^2+m_2^2+p^2){}^2-4 m_1^2 p^2}) \nonumber \\
   &\times \,_{2}F_{1}\left[\begin{array}{c}1,2-\frac{d}{2} ; \\ \frac{d}{2} ;\end{array} \frac{\left(-p^2+m_1^2-m_2^2+\sqrt{(p^2-m_1^2+m_2^2){}^2-4 p^2 m_1^2}\right){}^2}{4 p^2 m_2^2}\right]
\end{align}
The above relation can be viewed as a reduction formula without making reference to the underlying Feynman integral and the result is shown in Eq.\eqref{redbub1} and Eq. \eqref{redbub2}. In a similar manner, evaluation of other Feynman integrals can be used to obtain the relationship between hypergeometric functions \cite{Kniehl:2011ym}.   Such a reduction of hypergeometric functions with a higher number of variables to those with a lesser number of variables also helps when the analytic continuation has to be done to reach a certain kinematical region. For the case of Appell $F_4$ the elaborate analytic continuation has been performed explicitly in \cite{exton1995system} or using automatized algorithms \cite{Ananthanarayan:2021yar} for more general multi-variable hypergeometric functions. This whole process still does not guarantee convergence for all the values of the parameter space \cite{Bera:2022eag}. While for the case of $_2F_1$ complete table of analytic continuations is available \cite{becken2000analytic}. The procedure to find the analytic continuations also gets more complicated with the increase in the number of variables even with the use of automatized packages.\\
\section{\ar Package : Algorithm and Usage}\label{algorithm}
\subsection{Algorithm}
We now present a general algorithm for the case when we have $N$ denominators to find algebraic relation recursively.\\
Consider the general situation with product of $N$ denominators as $\dfrac{1}{d_{1} \cdots d_{N}}$. 
\begin{enumerate}
    \item We first find the algebraic relation by taking $d_{1}$ and $d_{2}$
\begin{equation}\label{decomp2deno}
    \frac{1}{d_{1}d_{2}} = \frac{x_{1}}{D_{1} d_{1}} + \frac{x_{2}}{D_{1} d_{2}}
\end{equation} 
\item We then multiply the above equation by $\dfrac{1}{d_{3}}$
\begin{equation}
    \frac{1}{d_{1}d_{2}d_{3}} = \frac{x_{1}}{D_{1} d_{1}d_{3}} + \frac{x_{2}}{D_{1} d_{2}d_{3}}
\end{equation}
\item We then find the algebraic relation of each pair of $d_{i}$s again using Eq.\eqref{decomp2deno}. 
\item Then in the resulting relation, we repeat this process until all the denominators are exhausted.
\end{enumerate}

The final result is a sum of $2^{N-1}$ terms where $N$ is the total number of denominators we started with. \\
It is to be noted that the above procedure is a slight modification of the original method\cite{Tarasov:2015wcd}. In \cite{Tarasov:2015wcd}, we start by seeking the following algebraic relation for the product of $N$ propagators
\begin{equation}\label{eq:algrelfull}
    \frac{1}{d_{1}\cdots d_{N}} = \frac{x_{1}}{D_{1}d_{1}\cdots d_{N-1}} + \cdots + \frac{x_{N}}{d_{2}\cdots d_{N}D_{1}}
\end{equation}
Comparing the coefficients of $k^{2},k$ and using the constant term we get an over-determined set of equations. Such a system leave $x_{3}, x_{4} \cdots x_{N}$ undetermined. Such procedure when used recursively with each term on the RHS of the above equation finally results in $N!$ total number of terms, unlike $2^{N-1}$ terms using the procedure presented here. Also, the arbitrariness in the choice of coefficients $x_{i}$s in the original algorithm is now present in the choice of parameters $M_{i}$s.   \\
\subsection{Usage}
The recursive algorithm presented previously has been automatized in the accompanying \mt package \ar. Below we demonstrate the usage of the package \ar.
After downloading the package and putting it in the same directory as the notebook we can call the package as follows: 
\begin{mmaCell}{Input}
SetDirectory[NotebookDirectory[]];
AlgRel.wl;
\end{mmaCell}
\begin{mmaCell}{Input}
<<AlgRel.wl
\end{mmaCell}
\begin{mmaCell}{Print}
AlgRel.wl v1.0
Authors : B. Ananthanarayan, Souvik Bera, Tanay Pathak
ROC2.wl v1.0
\end{mmaCell}

The package has been made assuming the form $d_{i}= (k+p_{i})^{2}-m_{i}^{2}$ for the propagator, where $k,p$ and $m$ can be changed as per the convenience of the user. The only command of the package is \ar, which can be called as follows
\begin{mmaCell}{Input}
AlgRel[\{Propagator's number\},\{k,q,m\},\{P,M\},x,Substitutions]
\end{mmaCell}
\begin{mmaCell}{Output}
\{\{Algebraic relation\},\{Values\}\}
\end{mmaCell}

The various elements of the input are as follows
\begin{itemize}
    \item \texttt{\{Propagator's number\}}: It is a list of numbers to denote various propagators. It need not necessarily be serial and to ease the use of the package in case of many propagators (See Section \ref{sec:doublebox} for an example).
    \item \texttt{\{k,q,m\}}: It is a list containing three variables corresponding to the general propagator $d_{i}=(k+q_{i})^{2}-m_{i}^{2}$. \texttt{k} denotes the loop momenta, \texttt{q} denotes the combination of external momenta and can be zero too and \texttt{m} denotes the mass of the propagator. 
    \item \texttt{\{P,M\}}: It is a list containing two variables. They are used to set the variables for the auxiliary propagator introduced for obtaining the algebraic relation, $D_{i}= (k+P_{i})^{2}-M_{i}^{2}$. It automatically takes the \texttt{k} from the previous list.
    \item \texttt{x}: It is used to denote the variable for the coefficients in the algebraic relation, Eq.\eqref{eq:algrelfull}.
   \item \texttt{Substitutions}: It is a list of substitution for $q_{i}$ and $M_{i}$.
\end{itemize}
The output of the above command is a nested list with two sub-lists with the following two sub-lists
\begin{itemize}
    \item \texttt{\{Algebraic relation\}}: It gives the algebraic relation for the product of propagators, Eq.\eqref{eq:algrelfull}.
    \item \texttt{\{Values\}}: It is a list of the values obtained for $\mathbf{P_{i}}$ and $x_{i}$.
\end{itemize}
Consider the example of Bubble integral. To obtain the result for it we can use the following command 

\begin{mmaCell}{Input}
AlgRel[\{1, 2\},\{k,q,m\},\{P, M\}, x,\{q[1]-> 0,q[2]->-p,M[1]->0\}]
\end{mmaCell}
\begin{mmaCell}{Output}
\{\{\mmaFrac{x[1]}{(\mmaSup{(k+P[1])}{2})(-\mmaSup{m[1]}{2}+\mmaSup{(k)}{2})}+\mmaFrac{x[2]}{(\mmaSup{(k+P[1])}{2})(-\mmaSup{m[2]}{2}+\mmaSup{(k-p)}{2})}\},

\{x[1]->\mmaFrac{\mmaSup{p}{2}+\mmaSup{m[1]}{2}-\mmaSup{m[2]}{2}+\mmaSqrt{\mmaSup{(\mmaSup{p}{2}+\mmaSup{m[1]}{2}- \mmaSup{m[2]}{2})}{2}-4\mmaSup{p}{2}(\mmaSup{m[1]}{2})}}{\mmaSup{p}{2}},...\}\}

\end{mmaCell}

Due to its length, the second element of the output (i.e., the substitution list) is not shown fully. It contain the values of the \texttt{x[1],x[2]} and \texttt{P[1]} as given in Eq.\eqref{bubxxp}. We remark that in the output all the vectors such as \texttt{P[1], q[1]} and \texttt{q[2]} will appear in bold. The scalar product \texttt{q[1].q[2]} appears \texttt{q[1]q[2]}. So care must be taken while doing the numerical checks for the same.

In the next section, we look at a few one-loop and two-loop examples where such a procedure is helpful. For cases where it was feasible to perform numerical checks using the integration, we perform them using \texttt{FIESTA5}\cite{Smirnov:2021rhf} and the corresponding results are given in appendix \ref{appendix:numerical}.
\section{Results}\label{results}
We now look at results for one loop and higher loop cases that are obtained with the help of the \ar package. All the results are also presented in the \mt file \texttt{Example.nb}.
\subsection{One-loop vertex integral}
\begin{figure}[htbp]
\centering\includegraphics[width=.3\textwidth]{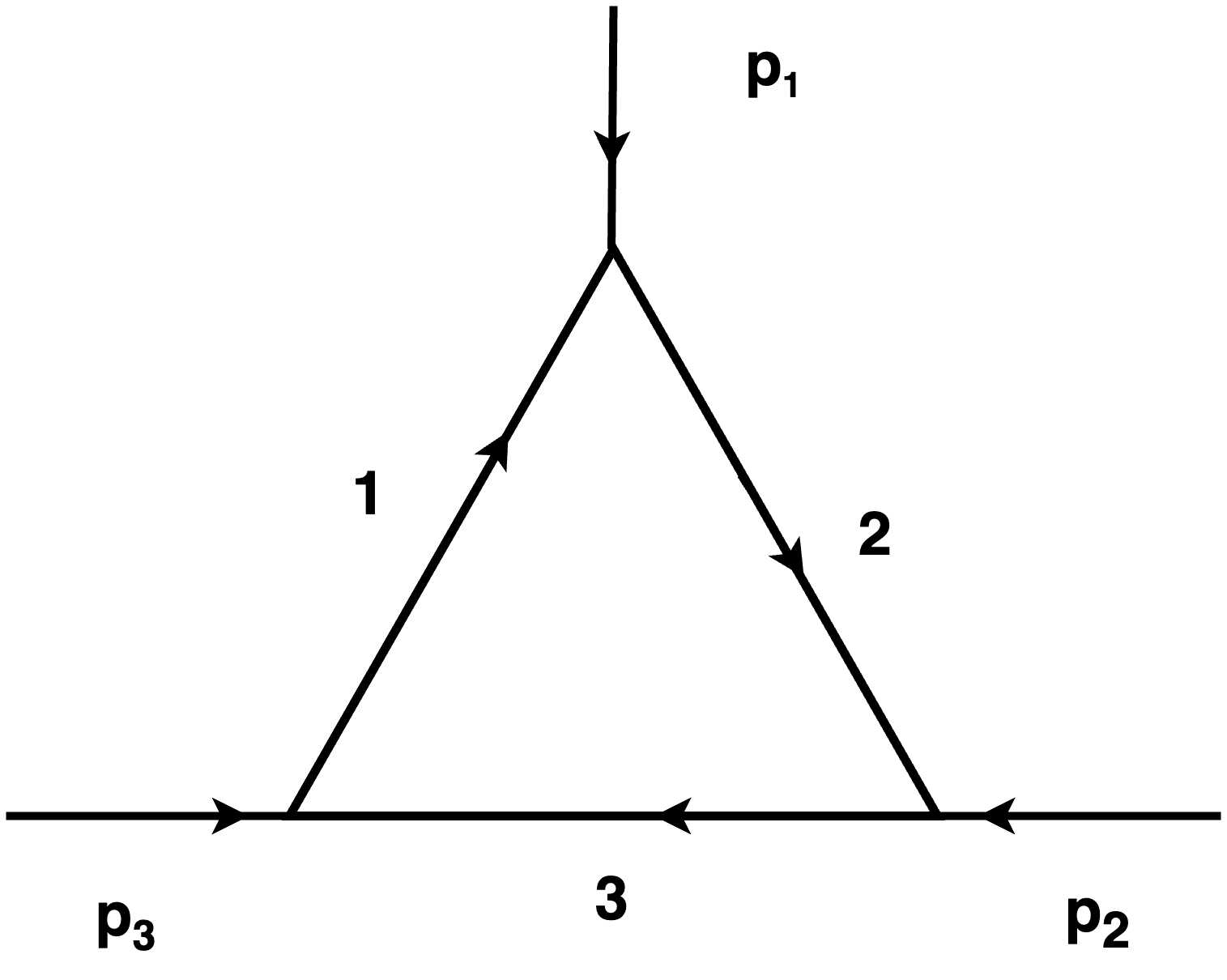}
\caption{Triangle diagram}\label{triangle}
\end{figure}
We consider the reduction of the one-loop vertex integral corresponding to Fig.\ref{triangle}, which is given by
\begin{equation}\label{vertexint}
I_{3}= \int \frac{d^{d}k}{(k^{2}-m_{1}^{2})((k+p_{1})^{2}-m_{2}^{2})((k+p_{1}+p_2)^{2} -m_{3}^{2})}
\end{equation}
We proceed as described in the previous section. We use the generalized propagators and do the substitutions accordingly so the result reduces to Eq.\eqref{vertexint}. This can be done using following command
\begin{mmaCell}{Input}
AlgRel[\{1,2,3\},\{k,q,m\},\{P,M\},x,\{q[1]->0,q[2]->p1,q[3]->p1+p2\}]
\end{mmaCell}
The result is a relation which is a sum of 4 terms, as follows
\begin{align}\label{vertexdecom}
 &\frac{x_1 x_3}{(k^2-m_1^2) (k+P_1){}^2 (k+P_2){}^2} +\frac{x_1 x_4}{(k+P_1){}^2 (k+P_2){}^2 ((k+p_1+p_2){}^2-m_3^2)} \nonumber \\ 
 &+\frac{x_2 x_5}{(k+P_1){}^2 (k+P_3){}^2 ((k+p_1){}^2-m_2^2)}
   +\frac{x_2 x_6}{(k+P_1){}^2 (k+P_3){}^2 ((k+p_1+p_2){}^2-m_3^2)}
\end{align}
where 
\begin{align*}
    x_1&= \frac{\sqrt{(m_1^2-m_2^2+p_1^2){}^2-4 m_1^2 p_1^2}+m_1^2-m_2^2+p_1^2}{2 p_1^2}, \nonumber \\
    x_2&= \frac{-\sqrt{(m_1^2-m_2^2+p_1^2){}^2-4 m_1^2 p_1^2}-m_1^2+m_2^2+p_1^2}{2 p_1^2}, \\
   \mathbf{P_1}&= \mathbf{p_1}-\frac{\mathbf{p_1} \left(-\sqrt{\left(m_1^2-m_2^2+p_1^2\right){}^2-4 m_1^2 p_1^2}-m_1^2+m_2^2+p_1^2\right)}{2 p_1^2},\nonumber \\
    x_3&=\frac{\sqrt{(m_1^2-m_3^2+(-p_1-p_2){}^2){}^2-4 m_1^2 (-p_1-p_2){}^2}+m_1^2-m_3^2+(p_1+p_2){}^2}{2 (-p_1-p_2){}^2},\\
    x_4&= \frac{-\sqrt{(m_1^2-m_3^2+(-p_1-p_2){}^2){}^2-4 m_1^2 (-p_1-p_2){}^2}-m_1^2+m_3^2+(p_1+p_2){}^2}{2 (-p_1-p_2){}^2},\nonumber \\
   \mathbf{ P_2}&= \frac{\left(\mathbf{-p_1-p_2}\right) \left(-\sqrt{\left(m_1^2-m_3^2+\left(p_1+p_2\right){}^2\right){}^2-4 m_1^2 \left(p_1+p_2\right){}^2}-m_1^2+m_3^2+\left(p_1+p_2\right){}^2\right)}{2 \left(p_1+p_2\right){}^2}+\left(\mathbf{p_1+p_2}\right) ,\\
    x_5&= \frac{\sqrt{(m_2^2-m_3^2+p_2^2){}^2-4 m_2^2 p_2^2}+m_2^2-m_3^2+p_1^2+(p_1+p_2){}^2-2 p_1 (p_1+p_2)}{2 p_2^2}, \nonumber \\
    x_6&= \frac{-\sqrt{(m_2^2-m_3^2+p_2^2){}^2-4 m_2^2 p_2^2}-m_2^2+m_3^2+p_1^2+(p_1+p_2){}^2-2 p_1 (p_1+p_2)}{2 p_2^2},\nonumber \\
    \mathbf{P_3}&= \left(\mathbf{p_1+p_2}\right)-\frac{\mathbf{p_2} \left(-\sqrt{-2 \left(m_2^2+m_3^2\right) p_2^2+\left(m_2^2-m_3^2\right){}^2+p_2^4}-m_2^2+m_3^2+p_2^2\right)}{2 p_2^2}
\end{align*}
Integrating Eq.\eqref{vertexdecom} over loop momenta $k$ we get vertex integral written as a sum of vertex integrals but with just one massive propagator. 

\subsection{One loop box integral}
\begin{figure}[htbp]
\centering\includegraphics[width=.25
\textwidth]{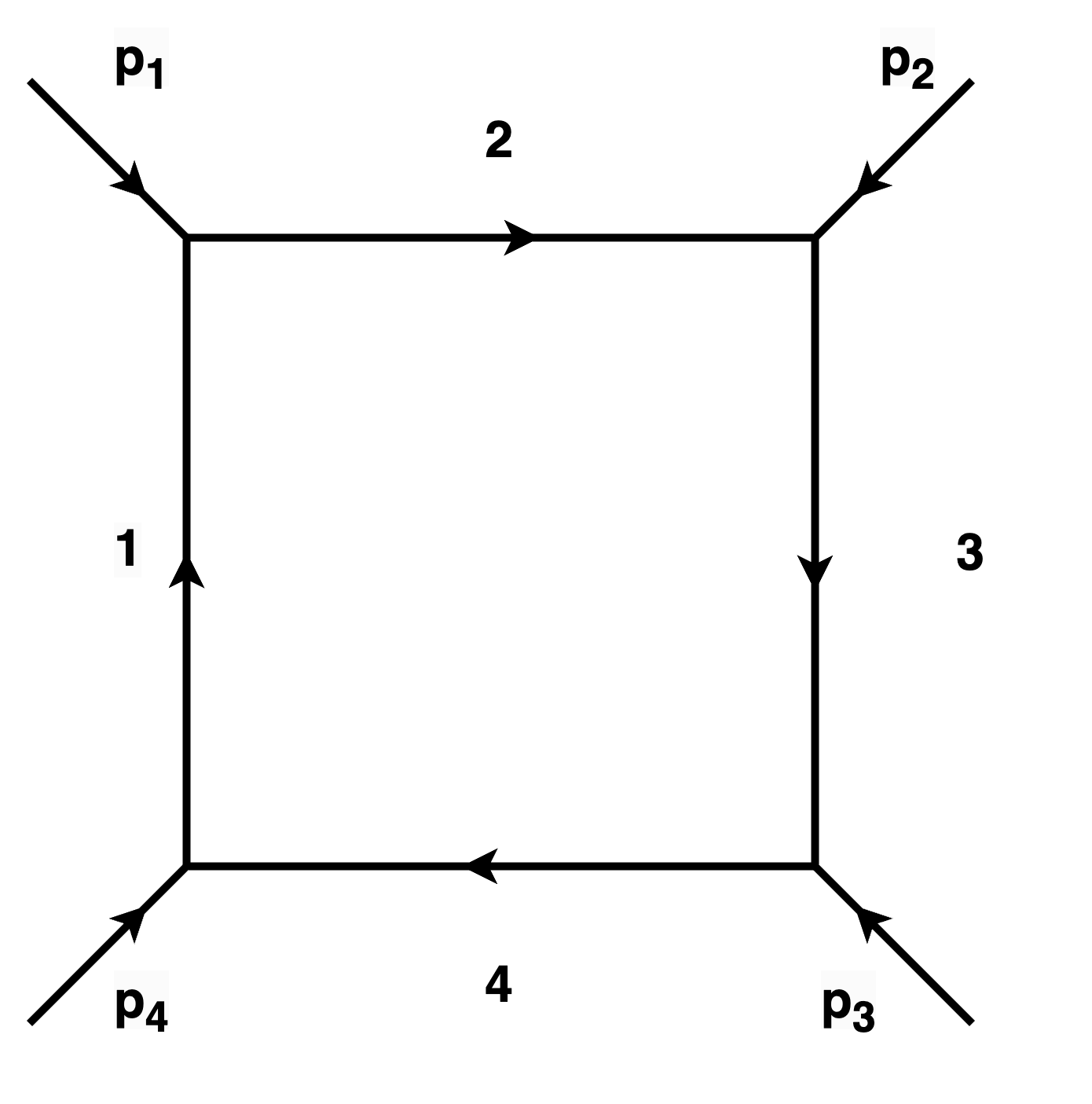}
\caption{Box diagram}\label{figbox}
\end{figure}

We now consider one loop box integral corresponding to Fig.\ref{figbox} which can be written as  

\begin{equation}
I_{4}= \int \frac{d^{d}k}{(k^{2}-m_{1}^{2})((k+p_{1})^{2}-m_{2}^{2})((k+ p_1 +p_2)^{2} -m_{3}^{2})((k+ p_1 +p_2+p_3)^{2} -m_{4}^{2})}
\end{equation}
We can get the algebraic relation using the following command 
\begin{mmaCell}{Input}
AlgRel[\{1,2,3,4\},\{k,q,m\},\{P,M\},x,\{q[1]->0,q[2]->p1,q[3]->p1+p2\\
,q[4]-> p1+p2+p3\}]

\end{mmaCell}
Substitute $q_{1}=0, q_2 = p_1,q_3= p_1+p_2,q_4= p_1+p_2+p_3$ and $M_{i}=0, i=1 \cdots 7 $ and simplifying we get 
\begin{align}\label{boxdecom}
&\frac{1}{(k^{2}-m_{1}^{2})((k+p_{2})^{2}-m_{2}^{2})((k+ p_2 +p_3)^{2} -m_{3}^{2})((k+ p_2 +p_3+p_4)^{2} -m_{3}^{2})} = \nonumber \\
&\frac{x_1 x_3 x_7}{(k^2-m_1^2) (k+P_1){}^2 (k+P_2){}^2 (k+P_4){}^2}+\frac{x_1 x_3 x_8}{(k+P_1){}^2 (k+P_2){}^2 (k+P_4){}^2 ((k+p_1+p_2+p_3){}^2-m_4^2)}\nonumber \\
&+\frac{x_1 x_4 x_9}{(k+P_1){}^2 (k+P_2){}^2 (k+P_5){}^2 ((k+p_1+p_2){}^2-m_3^2)}
\nonumber \\
&+\frac{x_1 x_4 x_{10}}{(k+P_1){}^2 (k+P_2){}^2 (k+P_5){}^2 ((k+p_1+p_2+p_3){}^2-m_4^2)} \nonumber \\
&+\frac{x_2 x_5 x_{11}}{(k+P_1){}^2 (k+P_3){}^2 (k+P_6){}^2 ((k+p_1){}^2-m_2^2)}\nonumber \\
&+\frac{x_2 x_5 x_{12}}{(k+P_1){}^2 (k+P_3){}^2 (k+P_6){}^2 ((k+p_1+p_2+p_3){}^2-m_4^2)} \nonumber \\
&+\frac{x_2 x_6 x_{13}}{(k+P_1){}^2 (k+P_3){}^2 (k+P_7){}^2 ((k+p_1+p_2){}^2-m_3^2)}+ \nonumber \\
&\frac{x_2 x_6 x_{14}}{(k+P_1){}^2 (k+P_3){}^2 (k+P_7){}^2 ((k+p_1+p_2+p_3){}^2-m_4^2)}
\end{align}
where the value of unknowns can be obtained from the \mt notebook \eg. Integrating Eq.\eqref{boxdecom} over loop momenta $k$ we get box integral written as a sum of 8 box integrals but with just one massive propagator. 
\subsection{One-loop pentagon integral}
\begin{figure}[htbp]
\centering\includegraphics[width=.25\textwidth]{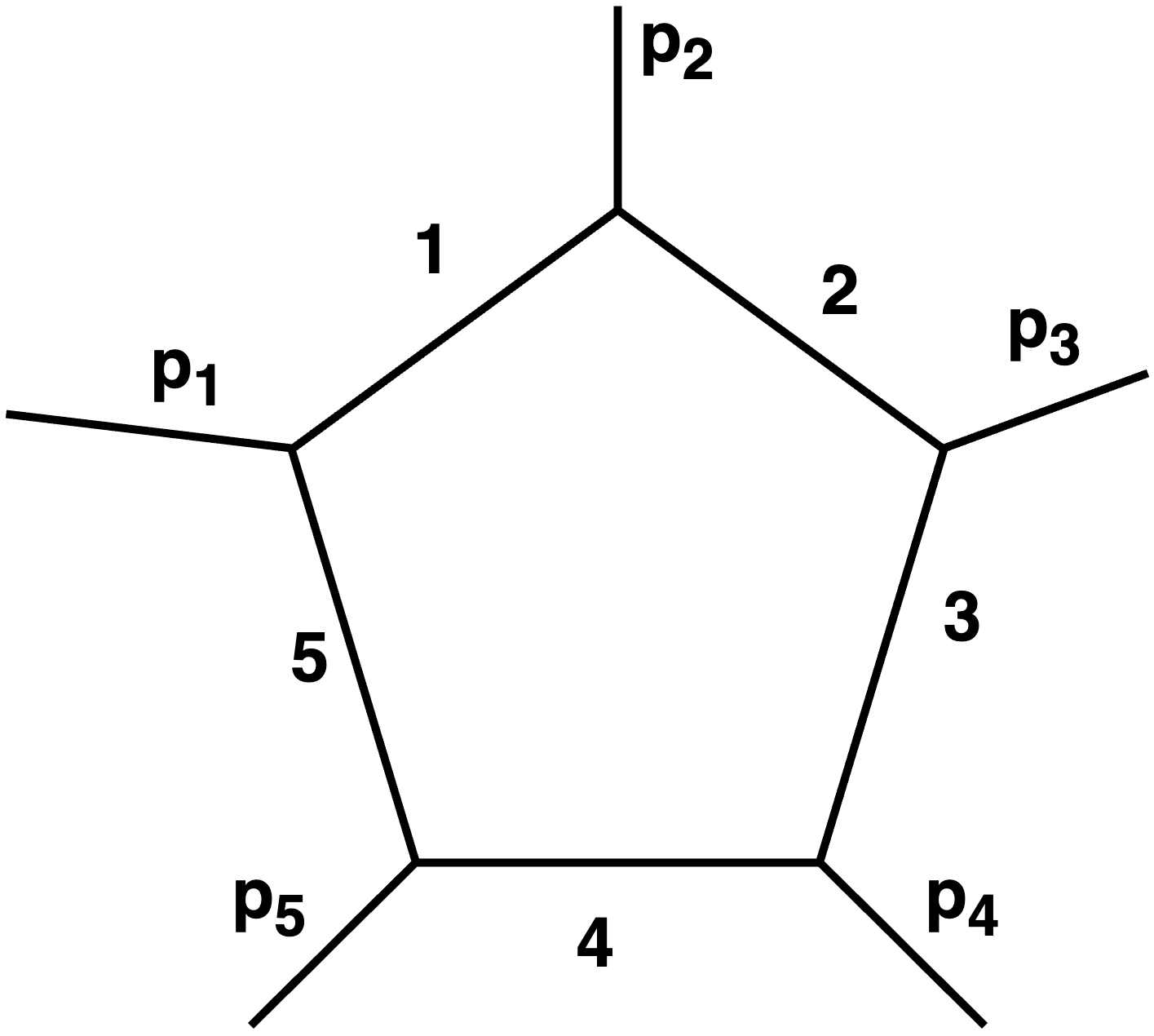}
\caption{Pentagon diagram}\label{pentagon}
\end{figure}
 The one-loop pentagon integral corresponding to Fig.\ref{pentagon} is given by \footnote{For this and the subsequent subsection we use the shorthand notation $p_{i_1}+p_{i_2} + \cdots = p_{i_{1} i_{2} \cdots}$, so as to avoid very lengthy expressions. }
 \begin{align}
I_{5}= \int \frac{d^{d}k}{(k^{2}-m_{1}^{2})((k+p_{1})^{2}-m_{2}^{2})((k+ p_{12})^{2} -m_{3}^{2})((k+ p_{123})^{2} -m_{4}^{2})((k+ p_{1234})^{2} -m_{5}^{2})}
\end{align}
We can get the algebraic relation using the following command 
\begin{mmaCell}{Input}
AlgRel[\{1,2,3,4,5\},\{k,q,m\},\{P,M\},x,\{q[1]->0,q[2]->p1,q[3]->p1+p2
,q[4]->p1+p2+p3,q[5]->p1+p2+p3+p4\}]

\end{mmaCell}
Doing the substitution as before and simplifying we get 
\begin{align}\label{pboxdecom}
&\frac{1}{(k^{2}-m_{1}^{2})((k+p_{1})^{2}-m_{2}^{2})((k+ p_{12})^{2} -m_{3}^{2})((k+ p_{123})^{2} -m_{4}^{2})((k+ p_{1234})^{2} -m_{5}^{2})} \nonumber \\
  &= \frac{x_1 x_3 x_7 x_{15}}{(k^2-m_1^2) (k+P_1){}^2 (k+P_2){}^2 (k+P_4){}^2 (k+P_8){}^2}\nonumber \\
  &+\frac{x_1 x_3 x_7 x_{16}}{(k+P_1){}^2 (k+P_2){}^2 (k+P_4){}^2 (k+P_8){}^2 ((k+p_{1234}){}^2-m_5^2)}\nonumber \\
   &+\frac{x_1 x_3 x_8 x_{17}}{(k+P_1){}^2 (k+P_2){}^2 (k+P_4){}^2 (k+P_9){}^2 ((k+p_{123}){}^2-m_4^2)}\nonumber \\
  &+\frac{x_1 x_3 x_8 x_{18}}{(k+P_1){}^2 (k+P_2){}^2 (k+P_4){}^2 (k+P_9){}^2 ((k+p_{1234}){}^2-m_5^2)}\nonumber \\
 & +\frac{x_1 x_4 x_9 x_{19}}{(k+P_1){}^2 (k+P_2){}^2 (k+P_5){}^2 (k+P_{10}){}^2 ((k+p_{12}){}^2-m_3^2)}\nonumber\\ 
 &+\frac{x_1 x_4 x_9 x_{20}}{(k+P_1){}^2 (k+P_2){}^2 (k+P_5){}^2 (k+P_{10}){}^2 ((k+p_{1234}){}^2-m_5^2)} \nonumber \\
   &+\frac{x_1 x_4 x_{10} x_{21}}{(k+P_1){}^2 (k+P_2){}^2 (k+P_5){}^2 (k+P_{11}){}^2 ((k+p_{123}){}^2-m_4^2)}\nonumber \\
   &+\frac{x_1 x_4 x_{10} x_{22}}{(k+P_1){}^2 (k+P_2){}^2 (k+P_5){}^2 (k+P_{11}){}^2 ((k+p_{1234}){}^2-m_5^2)} \nonumber \\
   &+\frac{x_2 x_5 x_{11} x_{23}}{(k+P_1){}^2 (k+P_3){}^2 (k+P_6){}^2 (k+P_{12}){}^2 ((k+p_1){}^2-m_2^2)}\nonumber \\
  &+\frac{x_2 x_5 x_{11} x_{24}}{(k+P_1){}^2 (k+P_3){}^2 (k+P_6){}^2 (k+P_{12}){}^2 ((k+p_{1234}){}^2-m_5^2)} \nonumber \\
   &+\frac{x_2 x_5 x_{12} x_{25}}{(k+P_1){}^2 (k+P_3){}^2 (k+P_6){}^2 (k+P_{13}){}^2 ((k+p_{123}){}^2-m_4^2)}\nonumber \\
   &+\frac{x_2 x_5 x_{12} x_{26}}{(k+P_1){}^2 (k+P_3){}^2 (k+P_6){}^2 (k+P_{13}){}^2 ((k+p_{1234}){}^2-m_5^2)} \nonumber \\
   &+\frac{x_2 x_6 x_{13} x_{27}}{(k+P_1){}^2 (k+P_3){}^2 (k+P_7){}^2 (k+P_{14}){}^2 ((k+p_{12}){}^2-m_3^2)}\nonumber \\
  &+\frac{x_2 x_6 x_{13} x_{28}}{(k+P_1){}^2 (k+P_3){}^2 (k+P_7){}^2 (k+P_{14}){}^2 ((k+p_{1234}){}^2-m_5^2)}\nonumber \\
   &+\frac{x_2 x_6 x_{14} x_{29}}{(k+P_1){}^2 (k+P_3){}^2 (k+P_7){}^2 (k+P_{15}){}^2 ((k+p_{123}){}^2-m_4^2)}+\nonumber \\
   &\frac{x_2 x_6 x_{14} x_{30}}{(k+P_1){}^2 (k+P_3){}^2 (k+P_7){}^2 (k+P_{15}){}^2 ((k+p_{1234}){}^2-m_5^2)}
\end{align}
where the values of $x_{i}$ and $P_{i}$ can be obtained from the \mt notebook \eg. Integrating Eq.\eqref{pboxdecom} over loop momenta $k$ we get pentagon integral written as a sum of 16 pentagon integrals but with just one massive propagator.
\subsection{One loop six-point integral}
\begin{figure}[htbp]
\centering\includegraphics[width=.25
\textwidth]{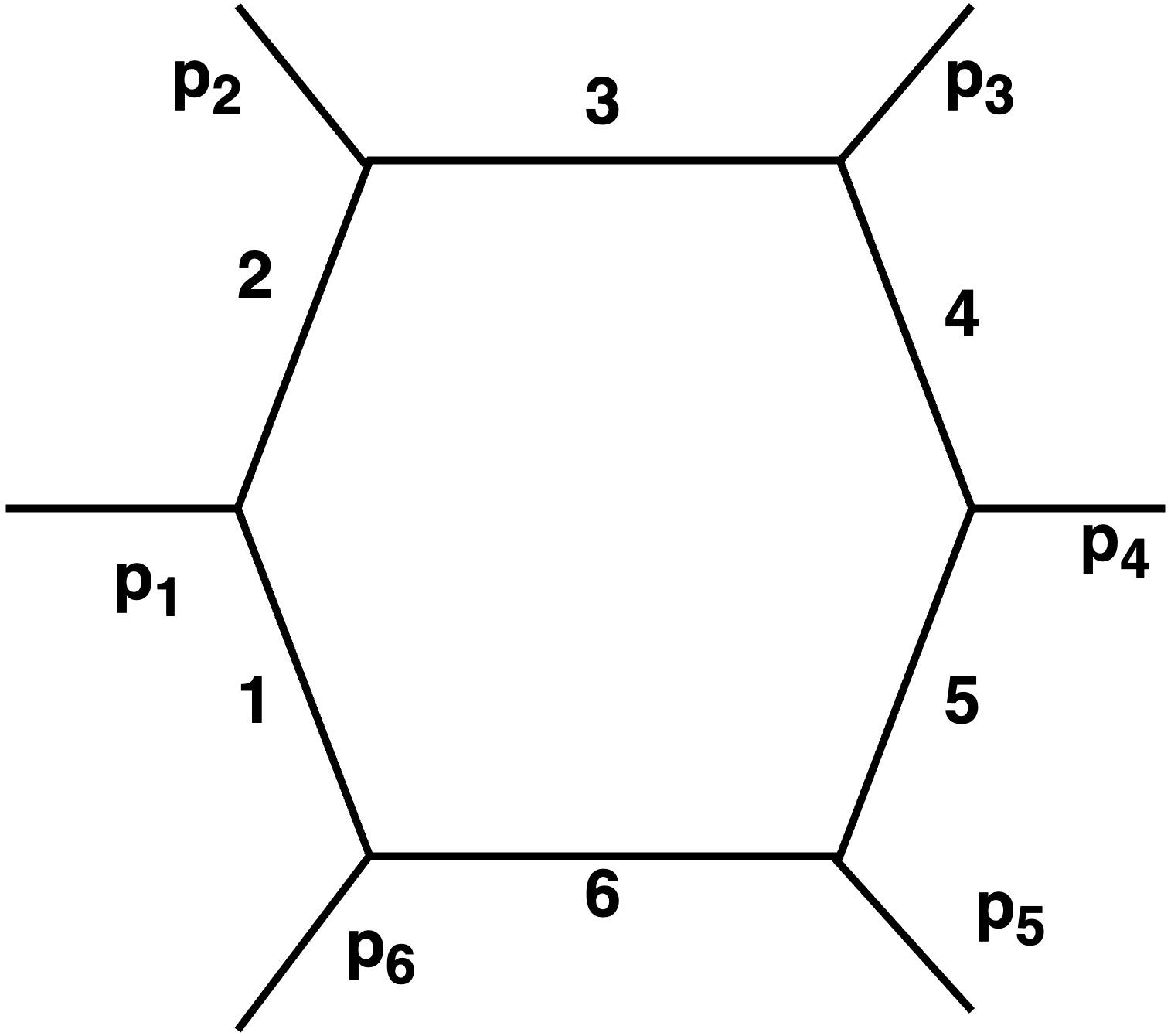}
\caption{Six point diagram}\label{hexagon}
\end{figure}
The six-point integral corresponding to the Fig.\ref{hexagon}, is
 \begin{align}
I_{6}= \int \frac{d^{d}k}{(k^{2}-m_{1}^{2})((k+p_{1})^{2}-m_{2}^{2})((k+ p_1 +p_2)^{2} -m_{3}^{2})((k+ p_{123})^{2} -m_{4}^{2})} \nonumber \\
\times \frac{1}{((k+ p_{1234})^{2} -m_{5}^{2})((k+ p_{12345} )^{2} -m_{6}^{2})}
\end{align}
As in the previous examples we use the following command to obtain the algebraic relations
\begin{mmaCell}{Input}
AlgRel[\{1,2,3,4,5,6\},\{k,q,m\},\{P,M\},x,\{q[1]->0,q[2]->p1,q[3]->p1+p2
,q[4]->p1+p2+p3,q[5]->p1+p2+p3+p4,q[6]->p1+p2+p3+p4+p5\}]

\end{mmaCell}
We omit the result as it is lengthy. The full result can be obtained from the \mt notebook \eg.

\subsection{Two-loop box integral}
\begin{figure}[htbp]
\centering\includegraphics[width=.25
\textwidth]{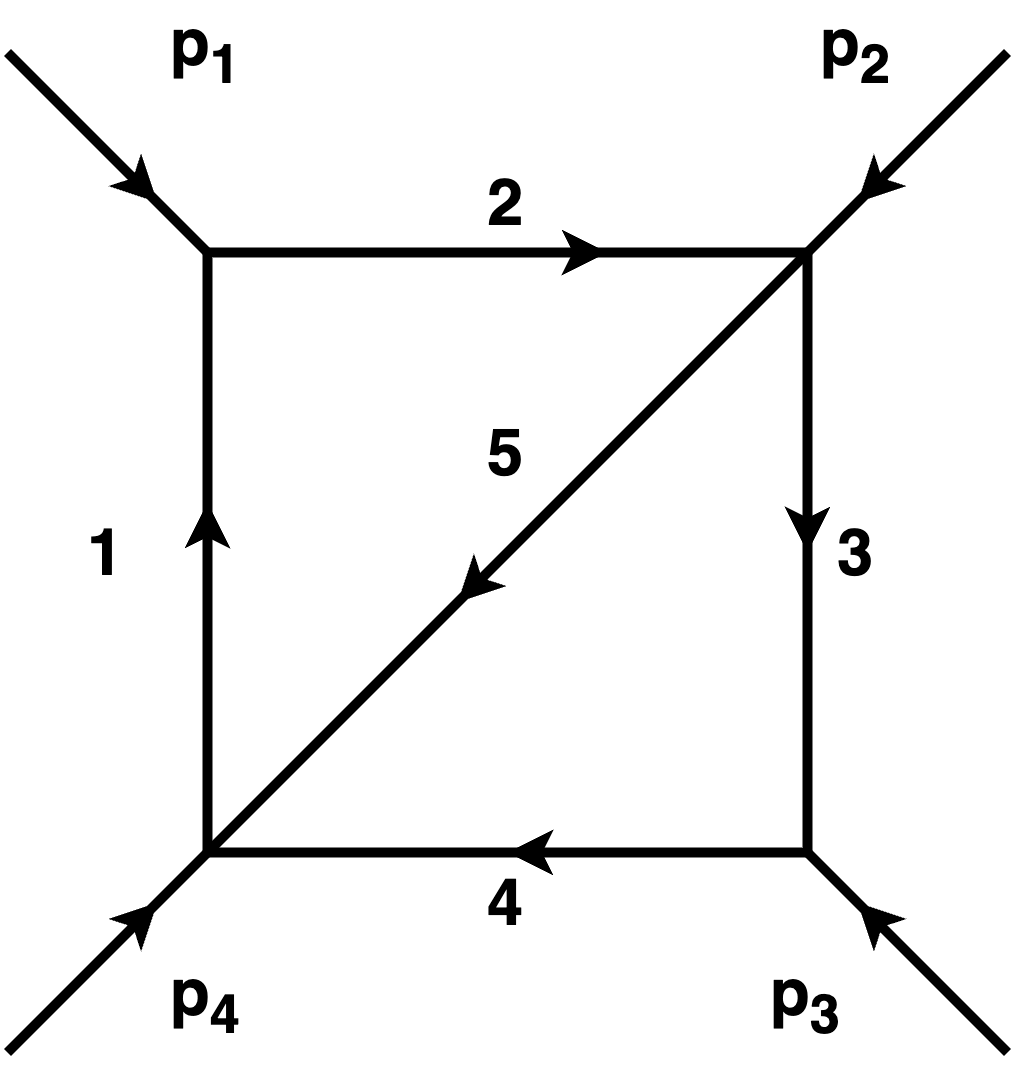}
\caption{Two loop Box diagram}\label{figtbox}
\end{figure}
To illustrate the method for higher loop integrals let us consider an example of the two-loop box integral, corresponding to the diagram Fig.\ref{figtbox}. The integral is as follows
\begin{equation}
I_{4,2}= \int \int \frac{d^{d}k_{1}d^{d}k_{2}}{(k_{1}^{2}-m_{1}^{2})((k_1+p_{1})^{2}-m_{2}^{2})(k_{2}^{2}-m_{3}^{2})((k_{2}+p_{3})^{2}-m_{4}^{2})((k_{1}-k_{2}+p_{1}+p_{2})^{2}-m_{5}^{2})}
\end{equation}
The propagators are numbered such that $i$ represents the propagator $d_{i}$.\\
Firstly we find the algebraic relation for the product of propagators numbered 1 and 2, which has only the loop-momenta $k_{1}$ we can use the following command
\begin{mmaCell}{Input}
AlgRel[\{1,2\},\{k1,q,m\},\{P,M\},x,\{q[1]->0,q[2]->p1\}]
\end{mmaCell}
Similarly, for propagators numbered 3 and 4 we can use the following command
\begin{mmaCell}{Input}
AlgRel[\{3,4\},\{k2,q,m\},\{Q,M\},y,\{q[3]->0,q[4]->p3\}]
\end{mmaCell}
The final relation that we obtain, with $M_{i}=0$ is( see \eg)
\begin{align}\label{decomtbox}
&\frac{1}{(k_{1}^{2}-m_{1}^{2})((k+p_{1})^{2}-m_{2}^{2})(k_{2}^{2}-m_{3}^{2})((k_{2}+p_{3})^{2}-m_{3}^{2})} = \frac{x_1 y_1}{(k_1^2-m_1^2) (k_2^2-m_3^2) (k_1+P_1){}^2 (k_2+Q_1){}^2}\nonumber \\
  &+\frac{x_2 y_1}{(k_2^2-m_3^2) (k_1+P_1){}^2 (k_2+Q_1){}^2 ((k_1+p_1){}^2-m_2^2)}+\frac{x_1 y_2}{(k_1^2-m_1^2) (k_1+P_1){}^2 (k_2+Q_1){}^2 ((k_2+p_3){}^2-m_4^2)} \nonumber \\
  &+\frac{x_2 y_2}{(k_1+P_1){}^2 (k_2+Q_1){}^2 ((k_1+p_1){}^2-m_2^2) ((k_2+p_3){}^2-m_4^2)}
\end{align}
where 
\begin{align}
x_1&= \frac{\sqrt{(m_1^2-m_2^2+p_1^2){}^2-4 m_1^2 p_1^2}+m_1^2-m_2^2+p_1^2}{2 p_1^2}, \nonumber\\
x_2&= \frac{-\sqrt{(m_1^2-m_2^2+p_1^2){}^2-4 m_1^2 p_1^2}-m_1^2+m_2^2+p_1^2}{2 p_1^2}\nonumber \\
\mathbf{P_1}&= \mathbf{p_1}-\frac{\mathbf{p_1} \left(-\sqrt{\left(m_1^2-m_2^2+p_1^2\right){}^2-4 m_1^2 p_1^2}-m_1^2+m_2^2+p_1^2\right)}{2 p_1^2}, \nonumber\\
y_{1}&=\frac{\sqrt{(m_3^2-m_4^2+p_3^2){}^2-4 m_3^2 p_3^2}+m_3^2-m_4^2+p_3^2}{2 p_3^2},\nonumber \\
y_2&= \frac{-\sqrt{(m_3^2-m_4^2+p_3^2){}^2-4 m_3^2 p_3^2}-m_3^2+m_4^2+p_3^2}{2 p_3^2},\nonumber\\
\mathbf{Q_1}&= \mathbf{p_3}-\frac{\mathbf{p_3} \left(-\sqrt{\left(m_3^2-m_4^2+p_3^2\right){}^2-4 m_3^2 p_3^2}-m_3^2+m_4^2+p_3^2\right)}{2 p_3^2}
\end{align}
Multiplying both sides of Eq.\eqref{decomtbox} by $\frac{1}{((k_{1}-k_{2}+p_{1}+p_{2})^{2}-m_{5}^{2})}$ gives the required algebraic relation for the two-loop box integral. 

\subsection{Two-loop double box integral}\label{sec:doublebox}
\begin{figure}[htbp]
\centering\includegraphics[width=.4
\textwidth]{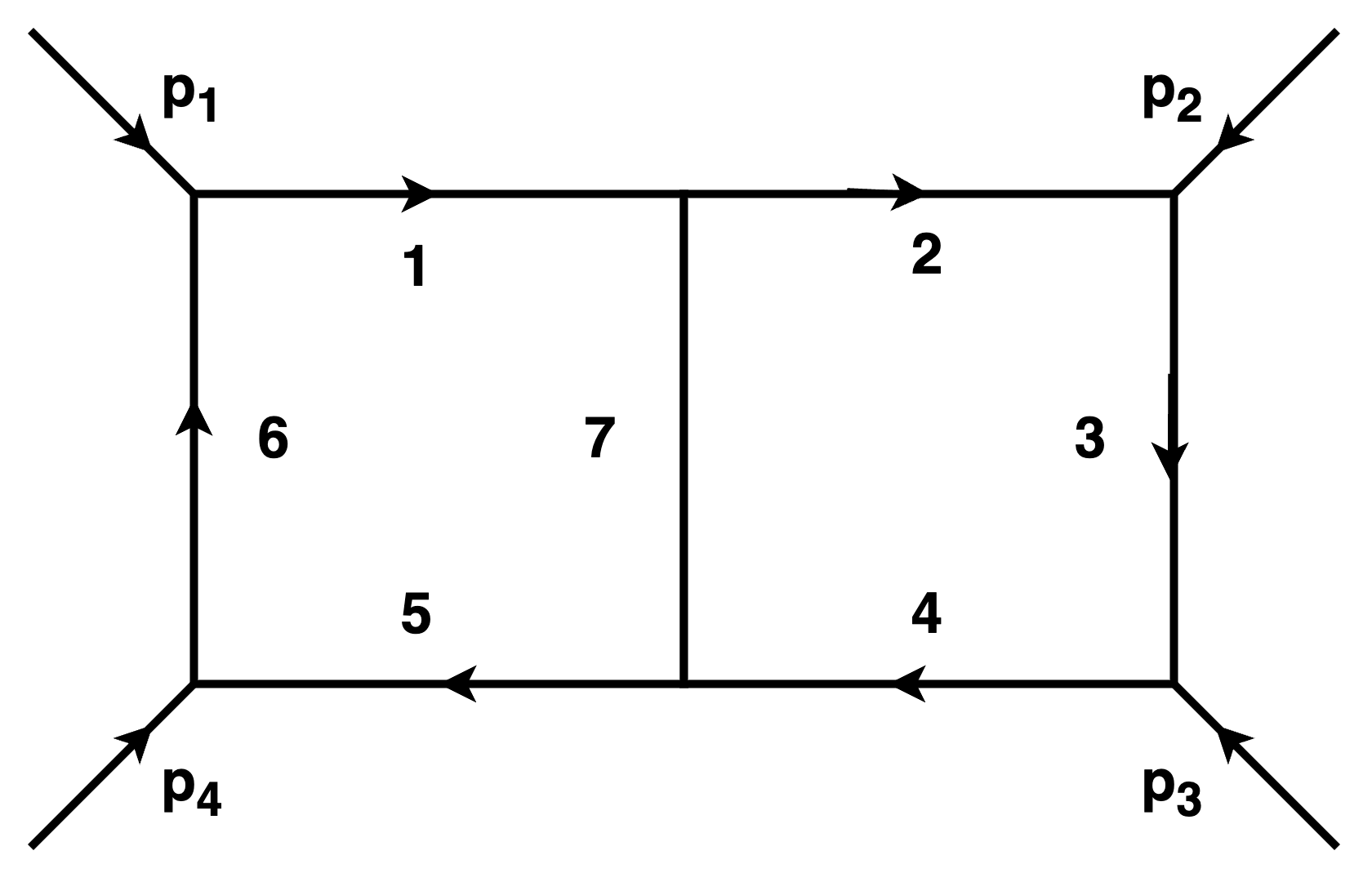}
\caption{Two loop Box diagram}\label{figdoublebox}
\end{figure}
Next, we consider the two-loop double-box integral corresponding to the diagram Fig.\ref{figdoublebox}. The integral is as follows
\begin{align}\label{dboxintegral}
I_{4,2}= \int \int \frac{d^{d}k_{1}d^{d}k_{2}}{(k_{1}^{2}-m_{1}^{2})(k_{2}^{2}-m_{2}^{2})((k_{2}+p_{2})^{2}-m_{3}^{2})((k_{2}+p_{23})^{2}-m_{4}^{2})((k_{1}+ p_{23})^{2}-m_{5}^{2})} \nonumber \\
\times \frac{1}{((k_{1}+ p_{234})^{2}-m_{6}^{2})((k_{1}-k_{2})^{2}-m_{7}^{2})}
\end{align}
The propagators are numbered such that $i$ represents the propagator $d_{i}$.\\
To find the algebraic relation for the product of propagators numbered 1,5 and 6 we can use the following command
\begin{mmaCell}{Input}
AlgRel[\{1,5,6\},\{k1,q,m\},\{P,M\},x,\{q[1]->0,q[5]->p2+p3,q[6]->p2+p3+p4\}]
\end{mmaCell}
Substituting value of $q_{1},q_{5}$ and $q_{6}$ corresponding to the Feynman integral we get
\begin{align}\label{algrel1}
    &\frac{x_1 x_4}{(k_1+P_1){}^2 (k_1+P_2){}^2 ((k_1+p_{234}){}^2-m_6^2)}+\frac{x_2 x_5}{(k_1+P_1){}^2 (k_1+P_3){}^2 ((k_1+p_{23}){}^2-m_5^2)} \nonumber \\ &+\frac{x_1 x_3}{(k_1^2-m_1^2) (k_1+P_1){}^2 (k_1+P_2){}^2}+\frac{x_2 x_6}{(k_1+P_1){}^2 (k_1+P_3){}^2 ((k_1+p_{234}){}^2-m_6^2)}
\end{align}
Similarly, for propagators numbered 2,3 and 4, we can use the following command
\begin{mmaCell}{Input}
AlgRel[\{2,3,4\},\{k2,q,m\},\{Q,M\},y,\{q[2]->0,q[3]->p2,q[4]->p2+p3\}]
\end{mmaCell}
which gives the following result after substituting the value of $q_{2},q_{3}$ and $q_{4}$ corresponding to the Feynman integral
\begin{align}\label{algrel2}
    &\frac{y_1 y_4}{\left(k_2+Q_1\right){}^2 \left(k_2+Q_2\right){}^2 \left(\left(k_2+p_{23}\right){}^2-m_4^2\right)}+\frac{y_1 y_3}{\left(k_2^2-m_2^2\right) \left(k_2+Q_1\right){}^2 \left(k_2+Q_2\right){}^2}\nonumber \\
   &+\frac{y_2 y_5}{\left(k_2+Q_1\right){}^2 \left(k_2+Q_3\right){}^2 \left(\left(k_2+p_2\right){}^2-m_3^2\right)}+\frac{y_2 y_6}{\left(k_2+Q_1\right){}^2 \left(k_2+Q_3\right){}^2 \left(\left(k_2+p_{23}\right){}^2-m_4^2\right)}
\end{align}
All the values of the parameters $P_{i}$,$Q_{i}$,$x_{i}$ and $y_{i}$ can be obtained from the \mt notebook \eg. To get the algebraic relation for the integrand in Eq.\eqref{dboxintegral} we multiply Eq.\eqref{algrel1} and \eqref{algrel2} together and then multiply both the sides of the equation by $\dfrac{1}{(k_{1}-k_{2})^{2}-m_{7}^{2}}$. 

We see that, unlike the one-loop case, we now have 3 massive propagators in each term. In fact with the present procedure to find the algebraic relation for any two-loop integral with all non-zero different masses, we have at least 3-massive propagators in each integral. Due to this reason, the present procedure won't be helpful for the case of integrals like the sunset integral where there are only 3-propagators.

\subsection{Three-loop ladder integral}
\begin{figure}[htbp]
\centering\includegraphics[width=.45
\textwidth]{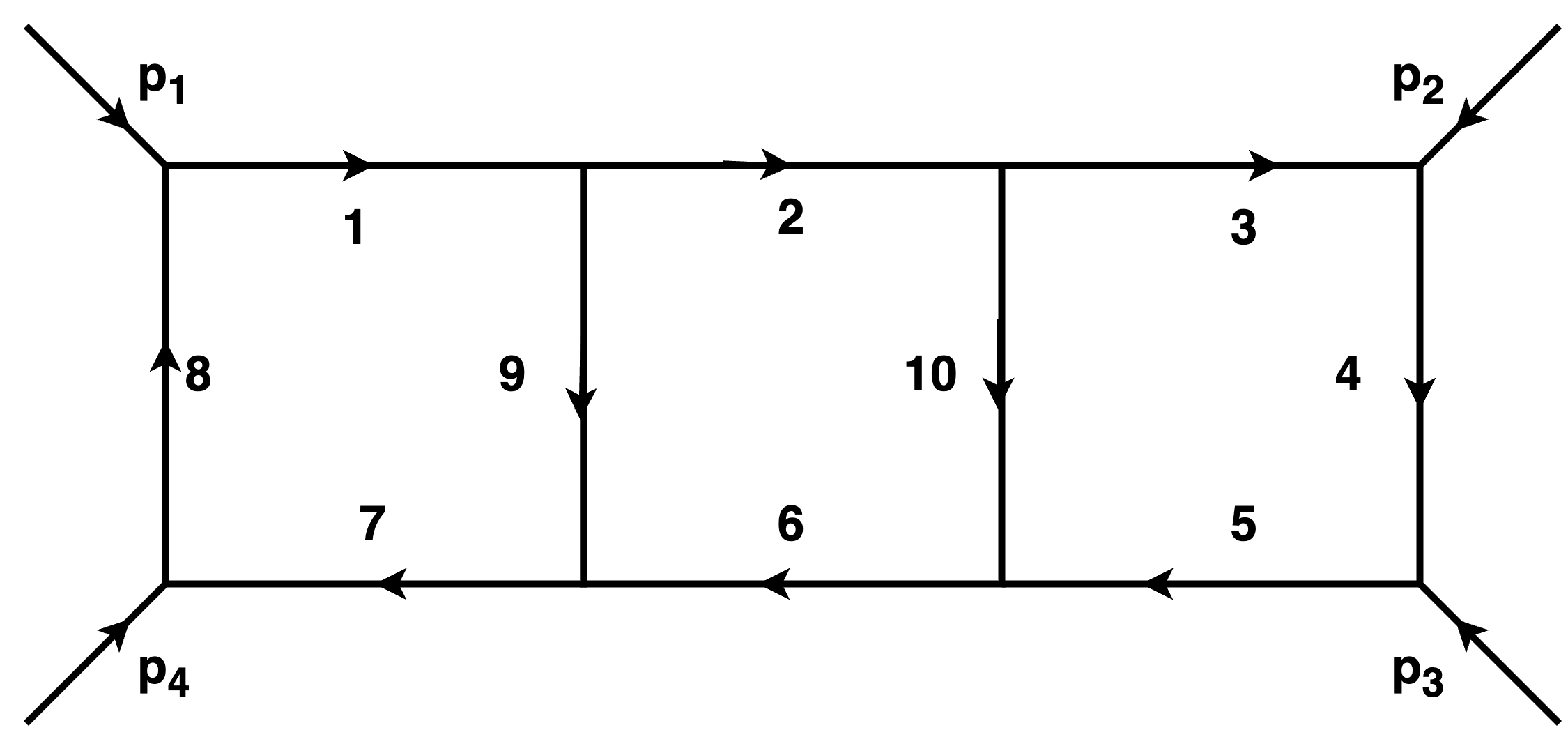}
\caption{Three-loop ladder diagram}\label{figtloopbox}
\end{figure}
The three-loop ladder integral corresponding to Fig.\ref{figtloopbox} is
\begin{align}
    I_{4,3}&= \int \int \int \frac{d^{d}k_{1}\,d^{d}k_{2}\,d^{d}k_{3}}{(k_{1}^{2}-m_{1}^{2})(k_{2}^{2}-m_{2}^{2})(k_{3}^{2}-m_{3}^{2})((k_{3}+p_{2})^{2}-m_{4}^{2})((k_{3}+p_{23})^{2}-m_{5}^{2})((k_{2}+ p_{23})^{2}-m_{6}^{2})} \nonumber \\
\times &\frac{1}{((k_{1}+ p_{23})^{2}-m_{7}^{2})((k_{1}+ p_{234})^{2}-m_{8}^{2})((k_{1}-k_{2})^{2}-m_{9}^{2})((k_{2}-k_{3})^{2}-m_{10}^{2})}
\end{align}
We use a similar strategy as before for this case too, to obtain the algebraic relation. The result contains 32 terms and is presented in the \mt file \eg.

\subsection{Limitation at higher loops}
Using sunset as an example we now demonstrate the limitation of the method for higher loop integrals. 
\begin{figure}[H]
\centering\includegraphics[width=.35\textwidth]{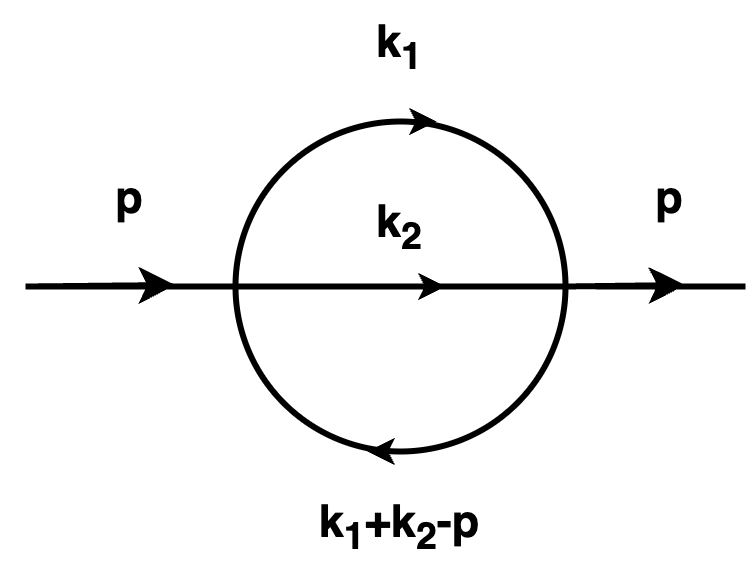}
\caption{Sunset diagram}\label{figsunset}
\end{figure}
As a demonstrative example, we consider the two-loop sunset integral as a starting point. The corresponding diagram is shown in Fig.\ref{figsunset}. The integral is given by 
\begin{equation}\label{sunsetint}
    I_{s}= \int \int \frac{d^{d}k_{1}d^{d}k_{2}}{(k_{1}^{2}-m_{1}^{2})(k_{2}^{2}-m_{2}^{2})((k_{1}-k_{2}+p)^2 - m_{3}^{2}))}
\end{equation}
Using the package we find the algebraic relation between the propagators $d_{1}$ and $d_{3}$. For this case, we obtain the following values of the coefficients $x_{1}$ and $x_{2}$. We further assume $M_{i}=0$
\begin{align}
    x_1= \frac{\sqrt{4 \left(p-k_2\right){}^2 \left(M_1^2-m_1^2\right)+\left(\left(p-k_2\right){}^2+m_1^2-m_3^2\right){}^2}-2 k_2 p+k_2^2+m_1^2-m_3^2+p^2}{2 \left(p-k_2\right){}^2} \nonumber \\
    x_2= \frac{-\sqrt{4 \left(p-k_2\right){}^2 \left(M_1^2-m_1^2\right)+\left(\left(p-k_2\right){}^2+m_1^2-m_3^2\right){}^2}-2 k_2 p+k_2^2-m_1^2+m_3^2+p^2}{2 \left(p-k_2\right){}^2}
\end{align}
We notice that for the case of sunset $q_{1}=0$ and $q_{3}= p-k_{2}$. Because of the fact that the coefficients $x_{1}$ and $x_{2}$ are dependent on loop momenta $k_{2}$, the final expression after finding the algebraic relation will include $x_{1}$ and $x_{2}$ in the integral. However, we can still consider the higher loops cases by taking only the propagators which are dependent on only single loop momenta together and propagators which are dependent on more than one loop momenta are excluded. 

\section{Reduction of Hypergeometric functions}\label{hypergeometric}

In this section, we study the examples when the Feynman integral evaluation gives results in terms of hypergeometric functions. The formalism to find algebraic relation for the product of propagators of Feynman integrals can be employed to find relations between hypergeometric functions \cite{inayat1987a,inayat1987b,Shpot:2007bz,Kniehl:2011ym}.
In this section, we point out some analytic results on $N-$point function \cite{davydychev1991some,davydychev1992general} and various hypergeometric relations that can be obtained from them with the present analysis.

It is well-known that the general one-loop $N-$point function with zero external momenta and different masses $(m_i, i = 1,\dots,N)$, with unit powers of propagators, can be expressed in terms of Lauricella  $F_D$ function \cite{davydychev1991some}

\begin{align}\label{npointgen1}
I^{(N)}(m_1,\dots, m_{N})&= \pi^{d / 2} i^{1-d}(-m_N^2)^{d / 2-N} \frac{\Gamma(N-d / 2)}{\Gamma(N)}  \nonumber\\
&\times F_D^{(N-1)}(N-\frac{d}{2}, 1, \ldots, 1 ; N \mid 1-\frac{m_1^2}{m_N^2}, \ldots, 1-\frac{m_{N-1}^2}{m_N^2})
\end{align}
where $F_{D}^{(L)}$ represents the Lauricella function of $L-$variables given by 
\begin{equation}
\begin{aligned}\label{lauricellafd}
F_D^{(L)} & (a, b_1, \ldots, b_L ; c \mid z_1, \ldots, z_L) = \sum_{j_1=0}^{\infty} \cdots \sum_{j_L=0}^{\infty} \frac{(a)_{j_1+\cdots+j_L}(b_1)_{j_1} \cdots(b_L)_{j_L}}{(c)_{j_1+\cdots+j_L}} \times \frac{z_1^{j_1} \cdots z_L^{j_L}}{j_{1} ! \cdots j_{L} !} .
\end{aligned}
\end{equation}
and $d$ is the dimension.  The general result (i.e., Eq. \eqref{npointgen1}) is a $N-1$ summation fold hypergeometric series. If one of the masses $m_{1},m_{2} \cdots, m_{N-1}$ vanishes then the function  $F^{(N-1)}_{D}$ reduces to $F^{(N-2)}_{D}$, using the following relation
\begin{align}\label{eq:fdlreduce}
F_D^{(L)}&\left(a, b_1, \ldots, b_{L-1}, b_L ; c \mid z_1, \ldots, z_{L-1}, 1\right) \nonumber \\
&= \frac{\Gamma(c) \Gamma\left(c-a-b_L\right)}{\Gamma(c-a) \Gamma\left(c-b_L\right)} \quad  F_D^{(L-1)}\left(a, b_1, \ldots, b_{L-1} ; c-b_L \mid z_1, \ldots, z_{L-1}\right)
\end{align}

Using the method presented, we can write the $N-$mass integral as a sum of integrals with just one mass, thus $N-1$ masses vanish. Then using Eq.\eqref{npointgen1} and \eqref{eq:fdlreduce} we can write each of these integrals as a term dependent only on mass $m_{i}$.  The whole result can be then expressed as a sum of terms each dependent on some $m_{i}$ . To evaluate the result of Eq. \eqref{npointgen1}, outside its associated region of convergence, one has to explicitly perform analytic continuation which is difficult to obtain at times for multi-variable hypergeometric functions. Such a reduction of the result is helpful when the analytic continuation of the Eq.\eqref{npointgen1} is required. Such a result should also be viewed as a  reduction formula for the Lauricella $F_{D}^{L}$, obtained using a physical problem \cite{Kniehl:2011ym} which can otherwise be hard to obtain.

For a general one-loop $N-$ point function with non-zero external momenta, the general result can be written as a generalized Lauricella hypergeometric function with $\frac{(N-1)(N+2)}{2}$ variables \cite{davydychev1992general}. For the case of general vertex integral, the result is a generalized Lauricella function with 5 variables \cite{davydychev1992general}. On the other hand, using Eq.\eqref{vertexdecom} the result can be written in terms of a hypergeometric function of $3-$variables \cite{davydychev1992general}.
Comparing Eq. \eqref{bub2f1} and \eqref{bubf4} we see that the evaluation of bubble integral has reduced from the evaluation of Appell $F_4$ which has two variables to that of hypergeometric $\,_2F_1$ with one variable. Such a result can be viewed as a general reduction formula without any explicit relation to the Feynman integrals it has been obtained from. Substituting $a=\frac{d}{2}, \frac{p^{2}}{m_{2}^{2}}=x$ and $\frac{m_{1}^{2}}{m_{2}^{2}}=y$,  we get the following relation
\begin{align}\label{redbub1}
 &y^{a-1}  F_{4}(a,1,a,a,x,y)-F_{4}(2-a,1,a,2-a,x,y) =\frac{1}{2 x}\left((-x+y-1)+\sqrt{-2 (x+1) y+(x-1)^2+y^2}\right) \nonumber \\ 
  &\times \,_{2}F_{1}\left[\begin{array}{c}1,2-a ; \\ a ;\end{array} \frac{\left(\sqrt{-2 (x+1) y+(x-1)^2+y^2}-x+y-1\right)^2}{4 x}\right]+((1-x-y)\nonumber\\
  &-\sqrt{-2 (x+1) y+(x-1)^2+y^2})\times y^{a-2}\,_{2}F_{1}\left[\begin{array}{c}1,2-a ; \\ a ;\end{array} \frac{\left(\sqrt{-2 (x+1) y+(x-1)^2+y^2}+x+y-1\right)^2}{4 x y}\right]\Bigg) 
\end{align}
Here a can take any value except negative integers and positive integers greater than 2.

We can further simplify the above relation by using the following relation of $F_{4}$ \cite{srivastava1985multiple}
\begin{align} 
&&F_4\left(\alpha, \beta ; \beta, \beta ;-\frac{x}{(1-x)(1-y)},-\frac{y}{(1-x)(1-y)}\right) =(1-x)^\alpha(1-y)^\alpha{ }_{2}F_{1}\left[\begin{array}{c}\alpha,\alpha-\beta+1  \\ \beta ;\end{array} xy\right].\\
&&
\end{align}
For our case $\alpha=1, \beta=a$. Thus we get
\begin{align}\label{redbub2}
&F_{4}(2-a,1,a,2-a,x,y) =\frac{1}{2 x}\left((-x+y-1)+\sqrt{-2 (x+1) y+(x-1)^2+y^2}\right) \nonumber \\ 
  &\times \,_{2}F_{1}\left[\begin{array}{c}1,2-a ; \\ a ;\end{array} \frac{\left(\sqrt{-2 (x+1) y+(x-1)^2+y^2}-x+y-1\right)^2}{4 x}\right]+((1-x-y)\nonumber\\
  &-\sqrt{-2 (x+1) y+(x-1)^2+y^2})\times y^{a-2}\,_{2}F_{1}\left[\begin{array}{c}1,2-a ; \\ a ;\end{array} \frac{\left(\sqrt{-2 (x+1) y+(x-1)^2+y^2}+x+y-1\right)^2}{4 x y}\right]\Bigg) \nonumber \\
&+ y^{a-1} \left(\frac{1-x-y-\sqrt{x^2-2 x (y+1)+(y-1)^2}}{2 x y}\right)  \,_{2}F_{1}\left[\begin{array}{c}1,2-a ; \\ a ;\end{array} \frac{(\sqrt{(x+y-1)^2-4 x y}+x+y-1)^2}{4 x y}\right]\nonumber \\
\end{align}
As a consequence of this we get $F_{4}(1,1;1,1;x,y)$ 
\begin{equation}
    F_{4}(1,1;1,1;x,y)= \frac{1}{\sqrt{(x+y-1)^2-4 x y}}
\end{equation}
We can also consider the result for the bubble integral with general masses and unit power of propagators, for which the result is given as follows \cite{davydychev1992general}
\begin{align}\label{bubkdf}
I_{2}= & (m_2^2)^{a-2} \Gamma(2-a) \times \sum_{j=0}^{\infty} \sum_{l=0}^{\infty} \frac{1}{j ! l !}(x)^j(1-y)^l  \times \frac{(2-a)_{j+l}(1)_{j+l}(1)_j}{(2)_{2 j+l}}
\end{align}
With the help of the reduction procedure, the result for bubble integral is given by Eq.\eqref{bub2f1}. This equality of Eq.\eqref{bubkdf} and \eqref{bub2f1} thus provides a reduction formula for the hypergeometric series in Eq.\eqref{bubkdf}, which can be written as follows
\begin{align}\label{kdft2f1}
 & \sum_{j=0}^{\infty} \sum_{l=0}^{\infty}    \frac{(2-a)_{j+l}(1)_{j+l}(1)_j}{(2)_{2 j+l}} \frac{(x)^j}{j!}\frac{(1-y)^l}{l!} =\frac{1}{2 x}\left((-x+y-1)+\sqrt{-2 (x+1) y+(x-1)^2+y^2}\right) \nonumber \\ 
  &\times \,_{2}F_{1}\left[\begin{array}{c}1,2-a ; \\ a ;\end{array} \frac{\left(\sqrt{-2 (x+1) y+(x-1)^2+y^2}-x+y-1\right)^2}{4 x}\right]+((1-x-y)\nonumber\\
  &-\sqrt{-2 (x+1) y+(x-1)^2+y^2})\times y^{a-2}\,_{2}F_{1}\left[\begin{array}{c}1,2-a ; \\ a ;\end{array} \frac{\left(\sqrt{-2 (x+1) y+(x-1)^2+y^2}+x+y-1\right)^2}{4 x y}\right]\Bigg)  
\end{align}

We can also obtain new hypergeometric relations by deriving other functional equations for the Feynman integrals(see appendix \ref{append:funcred}). Using Eq.\eqref{append:frbub2}, \eqref{bubequalmass} and  and we get 
\begin{align}
   & y^{a-1}  F_{4}(a,1;a,a;x,y)-F_{4}(2-a,1;a,2-a;x,y) = \nonumber \\  &\frac{(a-1)}{2 x}\Big( y^{a-2} (x+y-1)\, _2 F_1\left[\begin{array}{c}
1,2-a\\
\frac{3}{2}  ;\end{array} \frac{(x+y-1)^2}{4 x y} ;
\right] +
 (x-y+1)\, _2 F_1\left[\begin{array}{c}
1,2-a\\
\frac{3}{2}  ;\end{array} \frac{(x-y+1)^2}{4 x} ;
\right]  
\end{align}

We further obtain
\begin{align}\label{redbub3}
    &F_{4}(2-a,1;a,2-a;x,y) = \nonumber \\  &\frac{(a-1)}{2 x}\Big( y^{a-2} (x+y-1)\, _2 F_1\left[\begin{array}{c}
1,2-a\\
\frac{3}{2}  ;\end{array} \frac{(x+y-1)^2}{4 x y} ;
\right] +
 (x-y+1)\, _2 F_1\left[\begin{array}{c}
1,2-a\\
\frac{3}{2}  ;\end{array} \frac{(x-y+1)^2}{4 x} ;
\right]+ \nonumber \\
 &y^{a-1} (\frac{1-x-y-\sqrt{x^2-2 x (y+1)+(y-1)^2}}{2 x y})  \,_{2}F_{1}\Big[\begin{array}{c}1,2-a ; \\ a ;\end{array} \frac{(\sqrt{(x+y-1)^2-4 x y}+x+y-1)^2}{4 x y}\Big]
\end{align}
We provide a list of various reduction formulae that can be derived using Eq.\eqref{redbub1} and \eqref{redbub3} in the appendix \ref{appendix:redfor}.
The right-hand side of Eq.\eqref{redbub1} and \eqref{redbub3} can further be equated to give the relation between the sum of hypergeometric $_2F_1$ functions. An interesting consequence of this relation can be obtained with $a= \frac{3}{2}$
\begin{align}
    -\frac{\tanh ^{-1}(\frac{x-y+1}{2 \sqrt{x}})+\coth ^{-1}(\frac{2 \sqrt{x} \sqrt{y}}{x+y-1})}{2}&=\coth ^{-1}(\frac{2 \sqrt{x} \sqrt{y}}{\sqrt{x^2-2 x (y+1)+(y-1)^2}-x-y+1})- \nonumber \\
    &\tanh ^{-1}(\frac{\sqrt{-2 (x+1) y+(x-1)^2+y^2}+x-y+1}{2 \sqrt{x}})
\end{align}
As before, such a reduction also helps if the analytic continuation has to be performed to reach a certain kinematical region. We can find the analytic continuations for the series in Eq.\eqref{bubkdf} using automated tools \cite{Ananthanarayan:2021yar}, but it still does not guarantee that the parameter space has been covered. In contrast, the complete list of analytic continuations for the hypergeometric $_2F_1$\cite{becken2000analytic} is available and well implemented in software like \mt.  The complexity of the analytic continuation procedure also increases with the increase in the number of variables of the hypergeometric function due to the increase in difficulty to find the ROC of the resulting series.

We notice that the procedure is sufficiently general and one can obtain a large number of reduction formulae using it by doing the following steps
\begin{itemize}
    \item  We take the Eq.\eqref{npointgen1} or any other general result for $N-$point integral from \cite{davydychev1991some,davydychev1992general}.
    \item For a $N-$ point function we have a product of $N-$ propagators. We take any two propagators and find the algebraic relation. This results in a sum of 2 terms, for which the number of variables in the result, as in Eq.\eqref{npointgen1}, is reduced by one. This yields a reduction formula between, say $L$ (which is a function of $N$) variable hypergeometric function and $(L-1)$ variable hypergeometric function.
    \item We apply the previous step again, thus resulting in a relation between $L$ variable hypergeometric function and $(L-2)$ variable hypergeometric function. Also using the previous step it gives a relation between $(L-1)$ variable hypergeometric function and $(L-2)$ variable hypergeometric function.
    \item We apply the  procedure recursively  until we have an algebraic relation for the product of $N$ massive propagators as sum of $2^{N-1}$ terms, such that each term contains product of $N-$propagators with just one massive propagator.
    \item The final result of the procedure would be a collection of relations between $L, (L-1),...1$ variable hypergeometric functions. 
    
\end{itemize}

\section{Summary and Discussion}\label{conclusion}
We have presented an automatized package \ar for finding the algebraic relation for the product of propagators. These relations were used by Tarasov \cite{Tarasov:2015wcd,Tarasov:2008hw,Tarasov:2011zz,Tarasov:2022clb} to derive the Functional relations for Feynman integrals. The results obtained using the package are also sufficiently general and can be used further to obtain the functional relations for the Feynman integrals by appropriately choosing the arbitrary parameters. In the present work we focused on automatizing the method to derive algebraic relation for the propagators by suitably implementing a recursive algorithm (a slight modification to the Tarasov's algorithm\cite{Tarasov:2015wcd}). Furthermore, using a loop-by-loop approach we provided a systematic way so as to use these relation for higher loop integrals too. These relation occur with free parameters which can be chosen suitably.  Using various examples up to three-loops, we focused on how with a simple choice of these free parameters we can reduce integrals with large numbers of massive propagators into integrals with fewer massive propagators \cite{Tarasov:2011zz}, which can thus be computed easily. For the one-loop case, we obtained results for up to 6-point integral with the procedure and wrote them as a sum of $2^{N-1}$(for $N-$ point integral) integrals with one massive propagator.  We also showed how the procedure can be used for higher-loop integrals too where a loop-by-loop strategy has been applied for finding the relations.

Since the general results for the one-loop $N-$points integral are explicitly known for various cases in terms of multi-variable hypergeometric functions, we show how the present work can be used to obtain a large list of reduction formulae for these functions. As a demonstrative example of the same, we used the one-loop bubble integral where the reduction of the Appell $F_4$ to hypergeometric $_2F_1$ can be obtained. We also derive another reduction formula for a 2-variable hypergeometric series, Eq.\eqref{bubkdf} in terms of hypergeometric $_2F_1$. The relations thus obtained, can be treated as general reduction formulae for these functions without making reference to the Feynman integral they were derived from. These relation hence provides a way to derive non-trivial reduction formulae for multi-variable hypergeometric function using physical problems. They are also helpful, especially for situations where the analytic continuation of multi-variable hypergeometric functions has to be obtained to evaluate them outside their ROC, which is not easy to derive otherwise.

The present procedure of finding algebraic relation for the product of propagators can be used only if the propagators are dependent on just one loop-momenta. For this reason, the procedure cannot be applied with full generality to multi-loop integrals and a loop-by-loop approach has to be adopted. Hence the procedure is not helpful for integrals such as the sunset integral or in the cases where for each loop momenta $k_{i}$ there is just one propagator. To apply such a procedure to sunset-like integrals, a generalization of the procedure for the multi-variable case, when the propagators can depend on more than one loop momenta has to be developed. 

As we have seen that the algebraic relation obtained reduces the complexity of the Feynman integral. Specifically for the simple case of one loop bubble (in Section \ref{secMethod}), we saw that the result for general bubble integral, which was expressed in terms of double variable hypergeometric function Appell $F_4$, was reduced to $_2F_1$ which is a single variable hypergeometric function. It would be worth studying such reduction in complexity for other non-trivial cases of Feynman integrals which result in multi-variable hypergeometric functions of even higher variables. Since obtaining analytic expressions might not be feasible for such cases, a detailed numerical study for the same would be an important application of these algebraic relations after the proper function relations have been obtained by the proper choice of arbitrary variables. We would also like to point out to the possibility of using Lemma B.3., given in \cite{Flieger:2022xyq}, to find the similar relations as presented here.\footnote{We would like to thank William for pointing out this to us.}.

\appendix

\section{Functional reduction with $M_{i} \neq 0$}\label{append:funcred}
In this appendix, we point out other possibilities for the choice of arbitrary parameters $M_{i}$ \cite{Tarasov:2022clb}. This choice leads to different functional reduction equations than already presented. Also, this gives rise to different reduction formulae as has been done in section \ref{hypergeometric}.
Consider the bubble integral considered in section \ref{algorithm}. This time we choose a different non-zero value of $M_{1}$. Since the Feynman integrals are relatively easier to compute with equal masses a suitable choice is $M_{1}= m_{1}$. With this choice, we get, similar to Eq.\eqref{redeqbubble}, the following relation
\begin{equation}\label{redeqbubble2a}
    I_{2}(p^{2},m_{1},m_{2})= x_{1} I_{2}((P_1+p)^{2},m_1,m_{2})+ x_{2} I_{2}(P_{1}^{2},m_{1},m_1)
\end{equation}
with 
\begin{align}
    &x_1= \frac{m_1^2-m_2^2+p^2}{p^2}, \quad x_2= \frac{m_2^2-m_1^2}{p^2},\quad
    &\mathbf{P_1}= \frac{\mathbf{p} \left(m_{2}^2-m_{1}^2\right)}{p^2}-\mathbf{p}
\end{align}
We see that on the right-hand side of Eq.\eqref{redeqbubble2a}, we have a partial simplification. We can then exploit the symmetry of the $I_{2}$ integral under the exchange of $m_{1} \leftrightarrow m_{2}$. We do the exchange $m_{1} \leftrightarrow m_{2}$ in Eq.\eqref{redeqbubble2a} and add the resulting equation with it. Simplifying we get 
\begin{align}\label{append:frbub2}
   I_{2}(p^{2},m_{1},m_{2})=   \frac{p^{2}+m_{1}^{2}-m_{2}^{2}}{2p^{2}} I_{2}\left(\frac{\left(p^{2}+m_{1}^{2}-m_{2}^{2}\right)^{2}}{p^{2}},m_{1},m_{1}\right) +\nonumber \\  \frac{p^{2}-m_{1}^{2}+m_{2}^{2}}{2p^{2}} I_{2}\left(\left( \frac{p^{2}-m_{1}^{2}+m_{2}^{2}}{p}\right)^{2},m_{2},m_{2}\right)
\end{align}
The value of $I_{2}(p^{2},m,m)$ is \cite{Boos:1990rg}
\begin{equation}\label{bubequalmass}
I_{2}(p^{2},m,m)= m^{d-4} \Gamma\left(2-\frac{d}{2}\right){ }_2 F_1\left[\begin{array}{c}
1,2-\frac{d}{2}\\
\frac{3}{2}  ;\end{array} \frac{p^2}{4 m^2} ;
\right] 
\end{equation}
Substituting this in Eq.\eqref{append:frbub2} we get another functional equation for the bubble Feynman integral.
\section{Reduction formulae}\label{appendix:redfor}
\begin{align}
&F_{4}(1,1;1,1;x,y)=\frac{1}{\sqrt{(x+y-1)^2-4 x y}}\\
&F_{4}\left(\frac{3}{2},1;\frac{1}{2},\frac{3}{2};x,y\right)= \frac{x-y+1}{x^2-2 x (y+1)+(y-1)^2}\\
&F_{4}\left(\frac{5}{2},1;-\frac{1}{2},\frac{5}{2};x,y\right)= \frac{(x-y+1) \left(x^2-2 x (y+5)+(y-1)^2\right)}{\left(x^2-2 x (y+1)+(y-1)^2\right)^2}\\
 &F_{4}\left(\frac{1}{2},1;\frac{3}{2},\frac{1}{2};x,y\right)= \frac{\tanh ^{-1}\left(\frac{-\sqrt{-2 (x+1) y+(x-1)^2+y^2}+x-y+1}{2 \sqrt{x}}\right)}{\sqrt{x}} \\
&F_{4}\left(\frac{1}{2},1,\frac{3}{2};\frac{1}{2};x,y\right)=\frac{1}{2\sqrt{x}}(\tanh ^{-1}\left(\frac{x-y+1}{2 \sqrt{x}}\right)-2 \coth ^{-1}\left(\frac{2 \sqrt{x} \sqrt{y}}{\sqrt{(x+y-1)^2-4 x y}+x+y-1}\right)\nonumber \\
&+\coth ^{-1}\left(\frac{2 \sqrt{x} \sqrt{y}}{x+y-1}\right) )  \\
&F_{4}(0,1;2,0;x,y)= \frac{\sqrt{-2 (x+1) y+(x-1)^2+y^2}+x-y+1}{2 x} \nonumber \\
&F_{4}(2-a,1,a,2-a,1,1)= \frac{1}{2} \left(1-i \sqrt{3}\right) \, _{2}F_{1}\left[\begin{array}{c}1,2-a \\ a;\end{array} -\sqrt[3]{-1}\right]
\end{align}
\begin{align}
_{2}F_{1}\left[\begin{array}{c}1,2-a \\ \frac{3}{2};\end{array}\,\frac{(x-y+1)^2}{4 x}\right]&= \frac{\left(\sqrt{-2 (x+1) y+(x-1)^2+y^2}+(y-x-1)\right) }{(1-a) (x-y+1)}\nonumber \\
&_{2}F_{1}\left[\begin{array}{c}1,2-a \\ a;\end{array}\,\frac{\left((x-y+1)+\sqrt{(x-1)^2+y^2-2 (x+1) y}\right)^2}{4 x}\right]\\
_{2}F_{1}\left[\begin{array}{c}1,2-a \\ \frac{3}{2};\end{array}\frac{(x+y-1)^2}{4 x y}\right] &= \frac{\sqrt{-2 (x+1) y+(x-1)^2+y^2}-(x+y-1)}{(1-a) (x+y-1)}\nonumber \\
& _{2}F_{1}\left[\begin{array}{c}1,2-a \\ a;\end{array}\,\frac{\left(\sqrt{(x-1)^2+y^2-2 (x+1) y}-(x+y-1)\right)^2}{4 x y}\right]
\end{align}

We can use the following relation as given in \cite{srivastava1985multiple} and obtain formulae for $F_1$
\begin{equation}
\begin{aligned}
F_4 & {\left(\alpha, \beta ; \gamma, \beta ;-\frac{x}{(1-x)(1-y)},-\frac{y}{(1-x)(1-y)}\right) } =(1-x)^\alpha(1-y)^\alpha F_1(\alpha, \gamma-\beta, \alpha-\gamma+1 ; \gamma ; x, x y)
\end{aligned}
\end{equation}

We can further exploit the relation between $F_{1}$ and $F_{2}$ to derive reduction formulae for $F_{2}$. Wherever possible exploiting such relations amongst various hypergeometric functions we can derive reduction formulas for other hypergeometric functions.

In a similar manner using the following relation \cite{srivastava1985multiple} we can obtain formulae for $H_3$
\begin{equation}
\begin{aligned}
& F_4(\alpha, \beta ; \gamma, \beta ; x, y) =(1-x-y)^{-\alpha} H_3\left(\alpha, \gamma-\beta ; \gamma ; \frac{x y}{(x+y-1)^2}, \frac{x}{x+y-1}\right)
\end{aligned}
\end{equation}

\section{Numerical results}\label{appendix:numerical}
In this appendix, we present the results of the numerical checks using \texttt{FIESTA5} \cite{Smirnov:2021rhf}. We use the \ar package to obtain the algebraic relation and then perform the numerical integration of both the left-hand side, which is the original integral and the right-hand side which is the sum of integrals obtained using the algebraic relation. We give the results with 5 significant digits and $\mathcal{O}(\epsilon)$ in the Laurent expansion where $\epsilon= \frac{4-d}{2}$. 
\begin{enumerate}
    \item \textbf{Bubble integral}\\
    Parameters: $m_{1}^2= 1, m_{2}^2= 4, p^{2}=9$.\\
    LHS : $1.0758\, +3.61468 \epsilon+\frac{1.}{\epsilon}$.\\
    RHS :  $1.0758\, +3.61466 \epsilon+\frac{1.}{\epsilon}$.
    \item \textbf{Vertex integral}\\
    Parameters: $m_1^2= \frac{1}{2},m_2^2=\frac{1}{3},m_3^2= \frac{1}{5},p_1 p_2= 0,p_1^2= 4,p_2^2= 5$.\\
    LHS : $(-0.54587+0.13235 i)-(0.43590+1.4360 i) \epsilon$.\\
    RHS :  $(-0.54587+0.13235 i)-(0.43590+1.4360 i) \epsilon$.
    \item \textbf{Box integral}\\
    Parameters: $m_1^2= \frac{1}{10},m_2^2= \frac{2}{10},m_3^2= \frac{3}{10},m_4^2= \frac{5}{10},p_1 p_2= 0,p_1 p_3= 0,p_2 p_3= 0,p_1^2= 4,p_2^2= 5,p_3^2= 6$.\\
    LHS : $(0.06170+0.14437 i) \epsilon +(0.053719-0.010142 i)$.\\
    RHS :  $(0.06169+0.14437 i) \epsilon +(0.053719-0.010140 i)$.
    \item \textbf{Pentagon integral}\\
    Parameters: $m_1^2= \frac{1}{10},m_2^2=\frac{2}{10},m_3^2=\frac{3}{10},m_4^2=\frac{5}{10},m_5^2= \frac{6}{10},p_1 p_2= 0,p_1 p_3= 0,p_1 p_4= 0,p_2 p_3= 0,p_2 p_4= 0,p_3 p_4=0,p_1^2= 1,p_2^2= 2,p_3^2= 3,p_4^2= 4$.\\
    LHS : $(-0.082297+0.039722 i)-(0.24830+0.15504 i) \epsilon$.\\
    RHS :  $(-0.082298+0.039711 i)-(0.24830+0.15507 i) \epsilon$.
    \item \textbf{Two-loop box integral}\\
    Parameters: $m_1^2 =  \frac{1}{400},m_2^2 = \frac{9}{400},m_3^2 = \frac{1}{25},m_4^2 = \frac{1}{10},m_5^2 = \frac{1}{5},p_1^2 = 2,p_2^2 = 4,p_3^2 = 3,p_1 p_2 = 1,p_1 p_3 = 2,p_2 p_3 = 3$.\\
    LHS : $(-0.58821+0.79539 i)-(2.8199+3.9363 i) \epsilon$.\\
    RHS :  $(-0.58820+0.79537 i)-(2.8202+3.9361 i) \epsilon$.
\end{enumerate}
We find that the results obtained using the algebraic relations are numerically consistent. 
\bibliographystyle{jhep}
\bibliography{biblio.bib}
\end{document}